\documentclass[journal,comsoc]{IEEEtran}

\usepackage{graphicx}
\usepackage{amsmath,bbm,amssymb,amsfonts,amstext,amsopn}
\usepackage{cite}
\usepackage{balance}
\usepackage{url}
\usepackage{epsfig}
\usepackage{setspace}
\usepackage{stmaryrd}
\usepackage{psfrag}	
\usepackage{multirow}
\usepackage{float}
\usepackage[process=auto]{pstool}
\usepackage{etoolbox}
\usepackage{algorithm}
\usepackage{algorithmic}
\allowdisplaybreaks
\usepackage[nolist]{acronym}

\newcommand{\Nxtx}[1]{N_{#1}^{\mathrm{tx}}}
\newcommand{\Tsymb}{T^{\mathrm{symb}}}
\newcommand{\Ybxt}[2]{\bar{y}_{#1}\left(#2\right)}
\newcommand{\Ybxi}[2]{\bar{y}_{#1}^{(#2)}(\mathbf{s})}
\newcommand{\sML}{\hat{s}^{\mathrm{ml}}}
\newcommand{\Cxrt}[3]{C_{#1}(#2,#3)} 
\newcommand{\Gxrt}[3]{G_{#1}(#2,#3)}
\newcommand{\Cinxr}[2]{C_{#1}^{\mathrm{init}}(#2)}

\newtheorem{lem}{Lemma}
\newtheorem{remk}{Remark}

\newtheorem{prop}{Proposition}
\newtheorem{corol}{Corollary}

\newtoggle{arXiv}

\toggletrue{arXiv}

\iftoggle{arXiv}{%
  
}{%
}

\EndPreamble
\begin{document}
\title{Diffusive Molecular Communications with Reactive Molecules: Channel Modeling and Signal Design}
\author{
Vahid Jamali, \textit{Student Member IEEE}, Nariman Farsad, \textit{Member IEEE}, \\ Robert Schober, \textit{Fellow IEEE},  and Andrea Goldsmith, \textit{Fellow IEEE} \vspace{-1cm}

\thanks{This paper has been presented in part at IEEE ICC 2018 \cite{ICC_Reactive}.}
\thanks{This work was supported in part by the German Research Foundation under Project SCHO 831/7-1, in part by the Friedrich-Alexander
	University Erlangen-N{\"u}rnberg under the Emerging Fields Initiative, in part by the STAEDTLER Foundation, in part by the NSF Center for Science of Information under grant NSF-CCF-0939370, and in part by German Academic Exchange Service (DAAD).}
\thanks{V. Jamali and R. Schober are with the Institute for Digital Communications,
Friedrich-Alexander University (FAU), Erlangen D-91058, Germany (e-mail:
vahid.jamali@fau.de; robert.schober@fau.de).}
\thanks{N. Farsad and A. Goldsmith are with the Electrical Engineering Department,
Stanford University, Stanford, CA 94305 USA (e-mail: nfarsad@stanford.edu;
andrea@ee.stanford.edu).} 
}

\maketitle

\begin{acronym} 
\acro{MC}{molecular communication}
\acro{ISI}{inter-symbol interference}
\acro{ML}{maximum likelihood}
\acro{CR}{channel response}
\acro{PDE}{partial differential equation}
\acro{FDM}{finite difference method}
\acro{BER}{bit error rate}
\acro{RV}{random variable}
\acro{PMF}{probability mass function}
\acro{LLR}{log likelihood ratio}
\end{acronym}

\begin{abstract}
This paper focuses on \ac{MC} systems using two types of signaling molecules which may participate in a reversible bimolecular reaction in the channel. The motivation for studying these \ac{MC} systems is that they can realize the concept of constructive and destructive signal superposition, which leads to  favorable properties such as \ac{ISI} reduction and avoiding environmental contamination due to continuous release of signaling molecules into the channel.  This work first presents a general formulation for binary modulation schemes that employ two types of signaling molecules and proposes several modulation schemes as special cases. Moreover, two types of receivers are considered:  a receiver that is able to observe both types of signaling molecules (2TM), and a simpler receiver that can observe only one type of signaling molecules (1TM). For both of these receivers, the \ac{ML} detector for general binary modulation is derived under the assumption that the detector has perfect knowledge of the \ac{ISI}-causing sequence. The performance of this genie-aided \ac{ML} detector yields an upper bound on the performance of any practical detector. In addition, two suboptimal detectors of different complexity are proposed, namely an \ac{ML}-based detector that employs an estimate of the \ac{ISI}-causing sequence and a detector that neglects the effect of \ac{ISI}. The proposed detectors, except for the detector that neglects ISI for the 2TM receiver, require knowledge of the \ac{CR} of the considered \ac{MC} system. Moreover, the \ac{CR} is needed for performance evaluation of all proposed detectors.  However, deriving the \ac{CR} of \ac{MC} systems with reactive signaling molecules is challenging since the underlying partial differential equations that describe the reaction-diffusion mechanism are coupled and non-linear. Therefore, we develop an algorithm for efficient computation of the \ac{CR} and validate its accuracy via particle-based simulation. Simulation results  obtained using this \ac{CR} computation algorithm confirm the effectiveness of the proposed modulation and detection schemes. Moreover, these results show that \ac{MC} systems  with reactive signaling have superior performance relative to those with non-reactive signaling due to the reduction of \ac{ISI} enabled by the chemical~reactions. 
\end{abstract}


\acresetall
\section{Introduction}
Recent advances in biology, nanotechnology, and medicine have given rise to the need for communication in nano/micrometer scale dimensions \cite{Nariman_Survey,soldner2019survey}. In nature,  a common strategy for communication between nano/microscale entities such as bacteria, cells, and organelles (i.e., components of cells) is diffusive \ac{MC} \cite{soldner2019survey,CellBio}. In contrast to conventional wireless communication systems that encode information into electromagnetic waves, \ac{MC} systems embed information in the characteristics of  signaling molecules such as their type and concentration. Therefore,
diffusive \ac{MC} has been considered as a bio-inspired approach for communication between small-scale nodes for applications where conventional wireless communication may be inefficient or infeasible. In fact, over the last few years, several  testbeds have been developed as proof-of-concept for biological \cite{Nakano_Microplatform_2008,Akyildiz_testbed_2015,grebenstein_biological_2018} and non-biological \cite{Nariman_AcidBasePlatform,atthanayake2018experimental,unterweger_experimental_2018} diffusive \ac{MC} systems.

One characteristic of \ac{MC} is that the receiver always observes a \textit{constructive} superposition of the molecules  released in previous symbol intervals or by different transmitters. This feature leads to  several undesirable effects. First, many concepts in conventional communications that rely on both constructive and destructive superposition of signals, such as precoding, beamforming, and orthogonal sequences, are not  applicable in \ac{MC}. Second, the release of signaling molecules in consecutive symbol intervals introduces significant \ac{ISI} as the channel impulse response of \ac{MC} channels is heavy-tailed. Third, if molecules are continuously released, particularly into a bounded environment, the concentration of the signaling molecules increases over time and contaminates the environment. 

One solution to cope with these challenges is to use enzymes to degrade the signaling molecules in the environment \cite{Adam_Enzyme}. It has been shown in \cite{Adam_Enzyme} that \ac{ISI} is significantly reduced if enzymes are uniformly present in the channel. However, having uniformly distributed enzymes in the environment has two main drawbacks. First,   degradation of the signaling molecules via enzymes cannot be controlled, which may hurt performance.  Second, the \ac{ISI} reduction comes at the expense of  reducing the peak concentration of the signaling molecules observed at the receiver.  In \cite{Nariman_Acid}, the authors proposed to employ acids and bases for  signaling. This \ac{MC} system has the advantage that the release of the molecules can be controlled by the transmitter and acids and bases can react to cancel each other out. Hence, the contamination of the environment by signaling molecules is avoided. Moreover, the use of acids and bases implies the destructive and constructive superposition of signaling molecules \textit{in the channel} (not at the receiver) which can be exploited to reduce \ac{ISI}. In fact, the effectiveness of this reactive signaling for \ac{ISI} reduction  was \textit{experimentally} verified in \cite{Nariman_AcidBasePlatform}. These advantages of the \ac{MC} system in \cite{Nariman_Acid,Nariman_AcidBasePlatform} motivate us to consider \ac{MC} systems with reactive signaling molecules in this paper.

Knowledge of the \ac{CR} is typically needed for receiver design and performance evaluation. However,  for \ac{MC} with reactive signaling molecules, deriving the \ac{CR} is challenging since the underlying \acp{PDE} that describe the reaction-diffusion mechanism are \textit{coupled} and \textit{non-linear}. A closed-form solution to these equations has not yet been found, which has led to various approximations  \cite{Adam_Enzyme,NonlinearPDE_Debnath,PDE_numerical,ReactionDiffSim,Reza_Reaction}, see \cite{jamali2018channel} for a recent overview of various channel models that have been developed so far for \ac{MC} systems. For instance,  in \cite{Adam_Enzyme}, the spatial and temporal distribution of the enzyme concentration was assumed to be constant to obtain an approximate solution. However, for \ac{MC} systems in which the transmitter releases reactive signaling molecules into the channel, the concentrations of the molecules are temporally and spatially non-uniform and hence the constant distribution assumption does not hold. In the absence of closed-form solutions, numerical methods are commonly used to solve reaction-diffusion equations   in the chemistry and physics literature   \cite{PDE_numerical}.  This approach was applied to \acp{MC} in \cite{Nariman_Acid} where the authors employed a \ac{FDM} to solve the reaction-diffusion equation for a one-dimensional environment.  Another approach to compute the \textit{expected} concentrations of molecules is to average many realizations of concentrations obtained via a stochastic simulation  \cite{ReactionDiffSim,Chou_MasterEq,Adam_AcCoRD}. However, the computational complexity of these numerical and simulation methods is very high.  In \cite{Reza_Reaction}, data is encoded in the concentration difference of two types of molecules and it is shown that if identical diffusion coefficients for both types of signaling molecules are assumed then the resulting \ac{PDE}  for the concentration difference is linear. However, the statistical model for the difference of the observed molecules is still a function of the concentrations of both types of molecules.

In this paper, we consider an \ac{MC} system that employs two types of molecules for signaling where the signaling molecules may participate in a \textit{reversible bimolecular reaction}, such as the acid and base reaction in \cite{Nariman_Acid}. The considered reversible bimolecular reaction involves two reactions with different rates, namely the reaction of two reactant molecules that yields a product molecule and the decomposition of this product molecule into the two reactant molecules. 
 Moreover, we assume an unbounded (one-, two-, or three-dimensional\footnote{One- and two-dimensional environments are used as first-order approximations of three-dimensional environments in special cases to facilitate the analysis. For instance, propagation along long tubes can be approximated as one-dimensional \cite{InvGaussian} and propagation in a low-depth petri dish or thin tissues can be approximated as two-dimensional \cite{dy2008first}.}) environment and a passive receiver for simplicity. We first present a general formulation for binary modulation schemes that employ two types of signaling molecules and also propose several modulation schemes as special cases. Moreover, we consider two types of receivers, namely a receiver that is able to observe both types of signaling molecules (2TM) and a simpler receiver that can observe only one type of signaling molecules (1TM). For both types of receivers, we derive a genie-aided \ac{ML} detector that assumes perfect knowledge of the previously transmitted symbols, i.e., the \ac{ISI}-causing symbols. The performance of this genie-aided \ac{ML} detector yields an upper bound on the performance of any practical detector. In addition, we propose two suboptimal detectors of different complexity, namely an \ac{ML}-based detector that employs an estimate of \ac{ISI}-causing symbols and an \ac{ISI}-neglecting  detector that simply ignores the effect of \ac{ISI}. All of the proposed detectors, except the \ac{ISI}-neglecting detector for the 2TM receiver, require knowledge of the \ac{CR} of the considered \ac{MC} system. Moreover, the \ac{CR} is needed for performance evaluation of the proposed detectors. As discussed earlier, the computation of the \ac{CR} is complicated by the non-linearity that arises due to the bimolecular reaction and must be characterized for all possible sequences of numbers of molecules released by the transmitter into the channel. To address this issue, we develop an algorithm for efficient computation of the \ac{CR} of the considered \ac{MC} system for any arbitrary sequence of released numbers of molecules. This algorithm discretizes only the time variable and solves the resulting PDEs analytically in the space variables by decoupling the underlying reaction-diffusion equations and efficiently exploiting the simplifying characteristics inherent to the considered unbounded environment and passive receiver. This numerical method is much faster than algorithms that discretize both space and time, e.g., FDM used in \cite{Nariman_Acid}, and avoids their inherent numerical stability issues \cite{PDE_numerical}.  The accuracy of the proposed algorithm for \ac{CR} computation is validated using particle-based simulation.  Simulation results obtained using this \ac{CR} computation algorithm show that \ac{MC} systems  with reactive signaling have superior performance relative to those with non-reactive signaling due to the reduction of \ac{ISI} enabled by the chemical reactions. Moreover,  we show that unlike the \ac{MC} system in \cite{Adam_Enzyme},  \ac{ISI} is reduced in the considered \ac{MC} system without reducing the peak of the \ac{CR}. Furthermore, due to the considerable \ac{ISI} reduction by chemical reactions, the proposed suboptimal \ac{ML}-based detector performs very close to the genie-aided bound, and the proposed \ac{ISI}-neglecting detector causes only a small performance degradation with respect to the genie-aided bound.

We note that this paper expands its conference version \cite{ICC_Reactive} in several directions. First, only 2TM receivers were considered in \cite{ICC_Reactive}, whereas in this paper, we also consider 1TM receivers which are less complex. Moreover, in this paper, we derive a genie-aided detector and propose suboptimal detectors not only for 2TM receivers but also for 1TM receivers. Furthermore, we derive CR computation algorithms for one-, two-, and three-dimensional environments for completeness, whereas only a three-dimensional environment was considered in \cite{ICC_Reactive}. Finally, unlike \cite{ICC_Reactive}, in this paper, the concentration update rules of the proposed CR computation algorithm are further simplified for several special cases. 

The remainder of this paper is organized as follows. In Section~\ref{Sec:SysMod}, the considered system is presented.  The proposed optimal and suboptimal detectors are derived in Section~\ref{Sec:Analysis} and the proposed algorithm for computing the \ac{CR} is introduced in Section~\ref{Sec:Model}. Simulation results are reported in Section~\ref{Sec:SimResult}, and conclusions are drawn in Section~\ref{Sec:Conclusions}.

\section{System Model}\label{Sec:SysMod}

\begin{figure}
  \centering
 \scalebox{0.75}{
\pstool[width=1.2\linewidth]{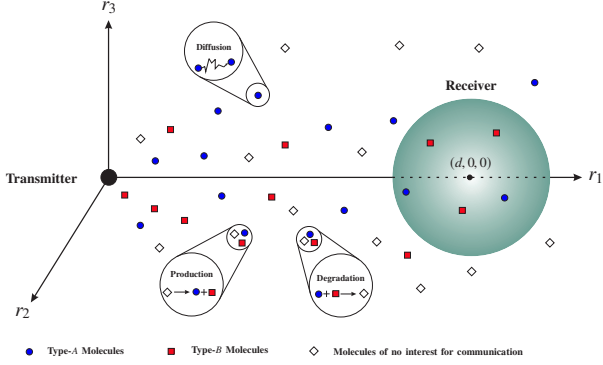}{
\psfrag{T}[l][c][0.7]{\textbf{Transmitter}}
\psfrag{R}[c][c][0.7]{\textbf{Receiver}}
\psfrag{D}[c][c][0.7]{$(d,0,0)$}
\psfrag{X}[c][c][1]{$r_1$}
\psfrag{Y}[c][c][1]{$r_2$}
\psfrag{Z}[c][c][1]{$r_3$}
\psfrag{Df}[c][c][0.45]{\textbf{Diffusion}}
\psfrag{Fl}[c][c][0.45]{\textbf{Flow}}
\psfrag{P}[c][c][0.45]{\textbf{Degradation}}
\psfrag{G}[c][c][0.45]{\textbf{Production}}
\psfrag{F1}[l][c][0.5]{\textbf{Type-$A$ Molecules}}
\psfrag{F2}[l][c][0.5]{\textbf{Type-$B$ Molecules}}
\psfrag{F3}[l][c][0.5]{\textbf{Molecules of no interest for  communication}}
} }   
\caption{Schematic illustration of the considered MC system with reactive signaling in a three-dimensional environment with coordinates $(r_1,r_2,r_3)$. }
\label{Fig:SysMod}
\end{figure}

The considered \ac{MC} system consists of a transmitter, a receiver, and a channel which are introduced in detail in the following. Fig.~\ref{Fig:SysMod} illustrates the considered system model in a three-dimensional environment. 

\subsection{Transmitter}\label{Sec:Transmitter}

\begin{figure}
	\centering
	\scalebox{0.6}{
		\pstool[width=1.5\linewidth]{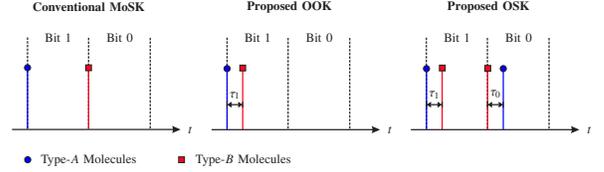}{
			\psfrag{M1}[c][c][0.8]{\textbf{Conventional MoSK}}
			\psfrag{M2}[c][c][0.8]{\textbf{Proposed OOK}}
			\psfrag{M3}[c][c][0.8]{\textbf{Proposed OSK}}
			\psfrag{a}[c][c][0.8]{Bit $1$}
			\psfrag{b}[c][c][0.8]{Bit $0$}
			\psfrag{c}[c][c][0.8]{$\tau_1$}
			\psfrag{d}[c][c][0.8]{$\tau_0$}
			\psfrag{t}[c][c][0.8]{$t$}
			\psfrag{E1}[l][c][0.8]{Type-$A$ Molecules}
			\psfrag{E2}[l][c][0.8]{Type-$B$ Molecules}
	} }   
	\caption{Schematic illustration of the considered binary modulation schemes, namely a) conventional MoSK, b) proposed OOK, and c) proposed OSK modulations. The vertical dashed lines represent the start of a new symbol interval and the vertical blue lines with a circle and the red lines with a square denote the release of type-$A$ and type-$B$ molecules, respectively.}
	\label{Fig:Modulation}
\end{figure}

We assume a point-source transmitter located at the origin of the $n$-dimensional Cartesian coordinate system, i.e., $\mathbf{r}=(r_1,\dots,r_n)=(0,\dots,0),\,\,n\in\{1,2,3\}$.  The transmitter employs two types of molecules for signaling, namely type-$A$ and type-$B$ molecules. In particular, the transmitter releases $\Nxtx{i}$ type-$i$ molecules into the channel at time instances $t\in\mathcal{T}_i,\,i\in\{A,B\}$. By properly defining $\mathcal{T}_i$, different modulation schemes can be realized. Let $s[k]\in\{0,1\}$ denote the binary symbol at the $k$-th symbol interval. In this paper, we focus on the following three binary modulation schemes: conventional molecule shift keying (MoSK) \cite{MoSK_Yilmaz,Arjmandi_isi_2017,Tepekule_isi_2015}, a proposed variant of on-off keying (OOK), and a new modulation scheme named order shift keying (OSK), see Fig.~\ref{Fig:Modulation}. 

\textit{Conventional MoSK:}
For binary one, $s[k]=1$, the transmitter releases $\Nxtx{A}$ type-$A$ molecules at the beginning of the symbol interval and no type-$B$ molecules whereas for binary zero, $s[k]=0$, the transmitter releases $\Nxtx{B}$ type-$B$ molecules at the beginning of the symbol interval and no type-$A$ molecules. For this modulation scheme, we obtain 
\begin{IEEEeqnarray}{lll} \label{Eq:Ti_MoSK}
\mathcal{T}_A=\left\{t|t=(k-1)\Tsymb,\,\,\forall k\,\,\mathrm{and}\,\, s[k]=1\right\} \IEEEyesnumber\IEEEyessubnumber \\
\mathcal{T}_B=\left\{t|t=(k-1)\Tsymb,\,\,\forall k\,\,\mathrm{and}\,\, s[k]=0\right\}, \IEEEyessubnumber
\end{IEEEeqnarray}
where $\Tsymb$ denotes the length of a symbol interval. 

\textit{Proposed OOK:} For binary one, $s[k]=1$, the transmitter releases $\Nxtx{A}$ type-$A$ molecules at the beginning of the symbol interval and $\Nxtx{B}$ type-$B$ molecules $\tau_1$ seconds after the start of the symbol interval. For binary zero, $s[k]=0$, the transmitter releases no molecules. Here, we choose $\tau_1$ as the peak of the \ac{CR} for the case where \textit{only} $\Nxtx{A}$ type-$A$ molecules are instantaneously released at $t=0$. For this modulation scheme, we obtain 
\begin{IEEEeqnarray}{lll} \label{Eq:Ti_OOK}
\mathcal{T}_A=\left\{t|t=(k-1)\Tsymb,\,\,\forall k\,\,\mathrm{and}\,\,s[k]=1\right\} \IEEEyesnumber\IEEEyessubnumber \\
\mathcal{T}_B=\left\{t|t=(k-1)\Tsymb +\tau_1,\,\,\forall k\,\,\mathrm{and}\,\,s[k]=1\right\}. \IEEEyessubnumber
\end{IEEEeqnarray}
Note that in contrast to conventional OOK modulation which employs only one type of molecule, here two types of molecules are used for  $s[k]=1$. The proposed modulation scheme effectively removes type-$A$ molecules that remain in the environment. As will be shown in Section~\ref{Sec:SimResult}, this modulation scheme  considerably reduces \ac{ISI} compared to conventional OOK modulation. A similar idea was introduced in \cite{Gau_AB} where the release patterns of type-$A$ and type-$B$ molecules were studied for efficient pulse shaping. However, the analysis in \cite{Gau_AB} relies on the linearity of the system and is not applicable to the  non-linear system considered in this paper.    

\textit{Proposed OSK:}
For binary one, $s[k]=1$, the transmitter releases $\Nxtx{A}$ type-$A$ molecules at the beginning of the symbol interval and $\Nxtx{B}$ type-$B$ molecules  $\tau_1$ seconds after the start of the symbol interval. In a similar manner, for binary zero, $s[k]=0$, the transmitter releases $\Nxtx{B}$ type-$B$ molecules at the beginning of the symbol interval and $\Nxtx{A}$ type-$A$ molecules  $\tau_0$ seconds after the start of the symbol interval. In particular, we choose $\tau_1$ ($\tau_0$) as the peak of the \ac{CR} assuming instantaneous release of \textit{only} $\Nxtx{A}$ ($\Nxtx{B}$) type-$A$ (type-$B$) molecules at $t=0$. For this modulation scheme, we obtain 
\begin{IEEEeqnarray}{lll} \label{Eq:Ti_PPM}
\mathcal{T}_A=\left\{t|t=(k-1)\Tsymb +(1-s[k])\tau_0,\,\,\forall k\right\} \IEEEyesnumber\IEEEyessubnumber \\
\mathcal{T}_B=\left\{t|t=(k-1)\Tsymb + s[k]\tau_1,\,\,\forall k\right\}. \IEEEyessubnumber
\end{IEEEeqnarray}
In other words, for OSK modulation, the information bit is encoded in the order in which the type-$A$ and type-$B$ molecules are released in each symbol interval.

To summarize, the basic idea behind the proposed  OOK and OSK modulation schemes is that after the release of a given type of signaling molecules (referred to as primary molecules), their reactive counterpart molecules (referred to as secondary/cleaning molecules) are released such that both types of molecules cancel each other out. This effectively shortens the \ac{CR} and reduces \ac{ISI}. The further the release times of both types of molecules are apart, the less efficient the cancellation and, as a result, the \ac{ISI} reduction become. However, if the two release times are too close together, the peak concentration of the primary molecules at the receiver will be severely reduced. To account for this trade-off while maintaining a simple design, we propose to release the secondary/cleaning molecules at the transmitter when the peak concentration of the primary molecules has already been observed at the receiver. To formalize this, let $\bar{c}_i(t),\,\,i\in\{A,B\}$, denote the expected concentration of type-$i$ molecules at the receiver  when only $N_i^{\mathrm{tx}}$ type-$i$ molecules are released at time $t=0$. Then, we choose $\tau_1=\mathrm{argmax}_t\,\bar{c}_A(t)$ and $\tau_0=\mathrm{argmax}_t\,\bar{c}_B(t)$.	

\subsection{Channel}\label{Sec:Channel}

We assume an unbounded $n$-dimensional environment for $n\in\{1,2,3\}$. 
The type-$A$ and type-$B$ molecules released by the transmitter diffuse in the environment with diffusion coefficients $D_A$ and $D_B$ in m$^2\cdot$s$^{-1}$, respectively, and may participate in the following bimolecular reaction
\begin{IEEEeqnarray}{lll} \label{Eq:Reaction}
A+B \underset{\kappa_b}{\overset{\kappa_f}{\rightleftharpoons}} \varnothing,
\end{IEEEeqnarray}
where $\kappa_f$ and $\kappa_b$ denote the forward reaction rate constant in molecule$^{-1}\cdot$m$^{n}\cdot$s$^{-1},\,\,n\in\{1,2,3\}$, and the backward reaction rate constant in molecule$\cdot$m$^{-n}\cdot$s$^{-1},\,\,n\in\{1,2,3\}$, respectively. Moreover, symbol $\varnothing$ denotes chemical species which are of no interest for communication. Note that (\ref{Eq:Reaction}) includes the reactions considered in \cite{Nariman_Acid,Reza_Reaction}. Moreover, if  type-$B$ molecules  represent enzymes and only type-$A$ molecules are used for signaling, (\ref{Eq:Reaction}) includes the degradation reaction in \cite{Adam_Enzyme} when the enzyme concentration is constant everywhere and is much larger than the concentration of the type-$A$ molecules such that the reaction in (\ref{Eq:Reaction}) does not change the enzyme concentration. 

\subsection{Receiver}\label{Sec:Receiver}

Let us assume that the receiver is located at distance $d$ from the transmitter. We consider two types of passive receivers: \textit{i)}  a receiver that is able to count the numbers of both type-$A$ and type-$B$ molecules within its volume, denoted by 2TM receiver; and \textit{ii)} a receiver that is able to count only type-$A$ molecules (1TM) within its volume, denoted by 1TM receiver. 1TM receivers are simpler to implement whereas 2TM receivers offer better performance since they exploit the diversity gain that observing two types of molecules provides. However, as is shown in Section~\ref{Sec:SimResult},  regardless of whether a 1TM or a 2TM receiver is employed, reactive \ac{MC} systems benefit from the ISI reduction enabled by employing two types of reactive signaling molecules.

 Let $\Ybxt{A}{t}$ and $\Ybxt{B}{t}$ denote the  \textit{expected} numbers of type-$A$ and type-$B$ molecules observed at the receiver at time $t$, respectively, due to the release of a known sequence of numbers of molecules by the transmitter. We refer to $\Ybxt{i}{t},\,\,i\in\{A,B\}$, as the \ac{CR} of the considered \ac{MC} system.  Note that depending on the length of the symbol interval, at a given time, the receiver may observe molecules released by the transmitter in multiple previous symbol intervals, i.e., \ac{ISI} may exist.  Let $L$ be the length of the channel memory\footnote{Theoretically, the memory length of the considered \ac{MC} channel is infinite; however, from a practical point-of-view, the effect of the previous symbols becomes negligible after several symbol intervals.} and $\mathbf{s}\in\{0,1\}^{L-1}$ denote the vector of the $L-1$ previously transmitted symbols. We also refer to $\mathbf{s}$ as the \ac{ISI}-causing sequence\footnote{For notational simplicity, we drop the symbol index $k$ in the remainder of the paper.}. Therefore, given $\mathbf{s}$ and $s$, the number of  type-$i$ molecules counted at the receiver at sample time $t_s$ is modelled as \cite{Yilmaz_Poiss,TCOM_MC_CSI}
\begin{IEEEeqnarray}{lll} \label{Eq:Poisson_ISI}
y_i \sim \mathcal{P}\left(s\Ybxi{i}{1}+(1-s)\Ybxi{i}{0}\right),\quad i\in\{A,B\},
\end{IEEEeqnarray}
where $\mathcal{P}(\lambda)$ denotes a Poisson \ac{RV} with mean $\lambda$. Moreover, $\Ybxi{i}{s}$ is $\Ybxt{i}{t_s}$ under the condition that the symbol in the current symbol interval is $s\in\{0,1\}$ and the \ac{ISI}-causing sequence is $\mathbf{s}\in\{0,1\}^{L-1}$. We note that the Poisson distribution assumed for $y_i$ in (\ref{Eq:Poisson_ISI}) is an approximation which has been shown to be accurate for reaction-diffusion processes in the chemistry and physics literature \cite{PoissonGardiner,CoxNatureCommun}. In Section~\ref{Sec:SimResult}, we will validate the Poisson model   in (\ref{Eq:Poisson_ISI})  using the particle-based simulator developed in \iftoggle{arXiv}{Appendix~F.}{\cite[Appendix~F]{Reactive_arXiv}.}

Note that due to the reaction process, the \ac{CR} of the considered \ac{MC} system $\Ybxi{i}{s}$ is a non-linear function of the transmitted data symbols $s$ and $\mathbf{s}$. Therefore, we cannot simply compute the \ac{CR} for one shot transmission and use convolution to capture the effect of  \ac{ISI} \cite{TCOM_MC_CSI}. 
In particular, to fully characterize the average behavior of the system, one has to compute the \ac{CR} for both symbol hypotheses $s\in\{0,1\}$ and all $2^{L-1}$ possible \ac{ISI}-causing sequences. In Section~\ref{Sec:Analysis}, we derive optimal and suboptimal detectors assuming $\Ybxi{i}{s},\,\,\forall s,\mathbf{s}$, is known. Then, in Section~\ref{Sec:Model}, we derive an efficient numerical algorithm for computation of \ac{CR} $\Ybxi{i}{s},\,\,\forall s,\mathbf{s}$.

\section{Detection Methods for Binary Modulation}\label{Sec:Analysis}

In this section, we derive the genie-aided \ac{ML} detector for binary modulation assuming the \ac{ISI}-causing sequence is known. This provides an upper bound on performance for any practical detector. Subsequently, we propose two suboptimal  practical detectors having different complexities.

\subsection{Optimal Genie-Aided ML Detector}\label{Sec:MLDetector}

In the following, we focus on symbol-by-symbol detection. We consider a genie-aided \ac{ML} detector that has  perfect knowledge of the \ac{ISI}-causing sequence.  

\textit{2TM Receivers:} For 2TM receivers, the genie-aided \ac{ML} detection problem for the considered binary modulation schemes is given by
\begin{IEEEeqnarray}{lll} \label{Eq:MLprobAB}
\sML &= \underset{s\in\{0,1\}}{\mathrm{argmax}}\,\,\Pr(y_A,y_B|s,\mathbf{s}) \nonumber \\
&\overset{(a)}{=} \underset{s\in\{0,1\}}{\mathrm{argmax}}\,\, f_{\mathcal{P}}(y_A|s,\mathbf{s})f_{\mathcal{P}}(y_B|s,\mathbf{s}),
\end{IEEEeqnarray}
where $\Pr(\cdot)$ denotes probability and $f_{\mathcal{P}}(x)=\frac{\lambda^x e^{-\lambda}}{x!}$ is the \ac{PMF} of a Poisson \ac{RV} with mean~$\lambda$. Equality $(a)$ follows from the fact that conditioned on $\Ybxi{i}{s},\,\,i\in\{A,B\}$, and $(s,\mathbf{s})$, RVs $y_A$ and $y_B$ are independent. 
The optimal detector for 2TM receivers is given in the following proposition.

\begin{prop}\label{Prop:MLDetector}
The genie-aided \ac{ML} detector for 2TM receivers as a solution of (\ref{Eq:MLprobAB}) is given by
\begin{IEEEeqnarray}{lll} \label{Eq:MLdetectorAB}
\sML = \begin{cases}
1, \quad &\mathrm{if}\,\, y_A \geq \alpha(\mathbf{s}) y_B + \beta(\mathbf{s}) \\
0, \quad &\mathrm{otherwise},
\end{cases}
\end{IEEEeqnarray}
where $\alpha(\mathbf{s})=\frac{1}{\chi(\mathbf{s})}\log\Big(\frac{\Ybxi{B}{0}}{\Ybxi{B}{1}}\Big)$, 
$\beta(\mathbf{s})=\frac{1}{\chi(\mathbf{s})}\big(\Ybxi{A}{1}+\Ybxi{B}{1}-\Ybxi{A}{0}-\Ybxi{B}{0}\big)$, and 
$\chi(\mathbf{s}) = \log\Big(\frac{\Ybxi{A}{1}}{\Ybxi{A}{0}}\Big)$.
\end{prop}
\begin{IEEEproof}
The proof is given in Appendix~\ref{App:Prop_MLDetector}.
\end{IEEEproof}

Proposition~\ref{Prop:MLDetector} reveals that the optimal detection rule for 2TM receivers involves only linear processing of the received signals $y_A$ and $y_B$ and a threshold comparison.  The coefficients of the corresponding linear operation, i.e., $\alpha(\mathbf{s})$ and $\beta(\mathbf{s})$, depend on the CR $\Ybxi{i}{s}$ and the ISI-causing sequence $\mathbf{s}$.

\textit{1TM Receivers:}  For 1TM receivers, the \ac{ML} problem is formulated as 
\begin{IEEEeqnarray}{lll} \label{Eq:MLprobA}
\sML &= \underset{s\in\{0,1\}}{\mathrm{argmax}}\,\,\Pr(y_A|s,\mathbf{s}) = \underset{s\in\{0,1\}}{\mathrm{argmax}}\,\, f_{\mathcal{P}}(y_A|s,\mathbf{s}).
\end{IEEEeqnarray}
The solution to the above problem is given in the following corollary.

\begin{corol}\label{Corol:MLDetector}
The genie-aided \ac{ML} detector as a solution of (\ref{Eq:MLprobA}) is given by
\begin{IEEEeqnarray}{lll} \label{Eq:MLdetectorA}
\sML = \begin{cases}
1, \quad &\mathrm{if}\,\, y_A \geq  \gamma(\mathbf{s}) \\
0, \quad &\mathrm{otherwise},
\end{cases}
\end{IEEEeqnarray}
where 
$\gamma(\mathbf{s})=\left(\Ybxi{A}{1}-\Ybxi{A}{0}\right)\big/\log\Big(\frac{\Ybxi{A}{1}}{\Ybxi{A}{0}}\Big)$.
\end{corol}
\begin{IEEEproof}
The proof is given in Appendix~\ref{App:Corol_MLDetector}.
\end{IEEEproof}

Corollary~\ref{Corol:MLDetector} shows that the optimal detection rule for 1TM receivers is a simple threshold detector.  The detection threshold, $\gamma(\mathbf{s})$, depends on the CR $\Ybxi{A}{s}$ and the ISI-causing sequence $\mathbf{s}$.

\begin{remk}
We note that the detectors in Proposition~\ref{Prop:MLDetector} and Corollary~\ref{Corol:MLDetector} are not limited to the specific modulation schemes introduced in Section~\ref{Sec:Transmitter} and can be applied in any (reactive or non-reactive) \ac{MC} system that employs binary modulation based on two types and one type of  signaling molecules, respectively. The adopted modulation scheme and whether or not the signaling molecules are reactive are reflected in the value of the \ac{CR}  $\Ybxi{i}{s}$, of course.
\end{remk}

The optimal detectors in Proposition~\ref{Prop:MLDetector} and Corollary~\ref{Corol:MLDetector} provide performance \textit{upper bounds} since they require  knowledge of the \ac{ISI}-causing sequence $\mathbf{s}$ which is not available in practice. Therefore, in the following, we propose some practical suboptimal detectors.

\subsection{Suboptimal Detectors}\label{Sec:Suboptimal}
In the following, we propose two categories of suboptimal detectors.

\subsubsection{ML-Based Detectors} Here, we assume that the receiver employs the detectors in Proposition~\ref{Prop:MLDetector} or Corollary~\ref{Corol:MLDetector} but uses its own estimates of the previous symbols as the \ac{ISI}-causing sequence.  This leads to suboptimal detectors which we refer to as ``\ac{ML}-based detectors with estimated \ac{ISI}". We show in Section~\ref{Sec:SimResult} that the performance of these \ac{ML}-based detectors with estimated \ac{ISI} is very close to the performance upper bound provided by the corresponding genie-aided \ac{ML} detector.

\subsubsection{ISI-Neglecting Detectors} Recall that the main motivation for considering \ac{MC} with reactive signaling as well as the proposed OOK and OSK modulation schemes was to reduce \ac{ISI}. In the following, we simplify the detectors given in Proposition~\ref{Prop:MLDetector} and Corollary~\ref{Corol:MLDetector} assuming an \ac{ISI}-free channel and propose to use them when the \ac{ISI} is sufficiently small. In particular, for an \ac{ISI}-free channel, parameters $\alpha(\mathbf{s})$,  $\beta(\mathbf{s})$, and $\gamma(\mathbf{s})$  do not depend on previous symbols and are denoted by $\alpha$, $\beta$, and $\gamma$, respectively.  This simplifies the detector in Proposition~\ref{Prop:MLDetector} for 2TM receivers to
\begin{IEEEeqnarray}{lll} 
\hat{s}= \begin{cases}
1, \quad &\mathrm{if}\,\, y_A \geq  y_B \\
0, \quad &\mathrm{otherwise},
\end{cases} \label{Eq:MLdetector_simpleAB}
\end{IEEEeqnarray}
and the detector in Corollary~\ref{Corol:MLDetector} for 1TM receivers to
\begin{IEEEeqnarray}{lll} 
\hat{s}= \begin{cases}
1, \quad &\mathrm{if}\,\, y_A \geq  \gamma \\
0, \quad &\mathrm{otherwise},
\end{cases} \label{Eq:MLdetector_simpleA}
\end{IEEEeqnarray}
where in (\ref{Eq:MLdetector_simpleAB}), we further assume symmetry with respect to the two molecule types, i.e., $D_A=D_B$ and $N_A=N_B$ hold\footnote{Note that if $D_A=D_B$ and $\Nxtx{A}=\Nxtx{B}$ hold, we obtain  $\Ybxi{A}{1}=\Ybxi{B}{0}\triangleq a$ and $\Ybxi{A}{0}=\Ybxi{B}{1}\triangleq b$. This leads to $\alpha=\frac{1}{\log(a/b)}\log(a/b)=1$ and $\beta=\frac{1}{\log(a/b)}(a+b-b-a)=0$ in Proposition~\ref{Prop:MLDetector}.}. 

\begin{remk}
Note that the detector in (\ref{Eq:MLdetector_simpleAB}) does not need knowledge of the \ac{CR} which makes it suitable for \ac{MC} systems with limited computational capabilities. However, for the detector in (\ref{Eq:MLdetector_simpleA}), the computation of $\gamma$ requires knowledge of the \ac{CR}, see e.g. \cite[Eq.~(5)]{TCOM_NonCoherent} for the optimal value of $\gamma$ for an \ac{ISI}-free \ac{MC} channel.  On the other hand, the advantage of the detector in (\ref{Eq:MLdetector_simpleA}) over that in (\ref{Eq:MLdetector_simpleAB}) is that the corresponding receiver is simpler because it needs to  be able to count only one type of molecule.
\end{remk}

\section{Computation of Channel Response   \\ for MC Systems with Reactive Signaling Molecules}\label{Sec:Model}
In this section, we first formally present the problem statement for \ac{CR} computation. Next, we derive a numerical algorithm for computing the \ac{CR} and discuss its complexity with respect to the available numerical methods for \ac{CR} computation.

\subsection{Problem Statement}\label{Sec:Problem}
Let $\Cxrt{A}{\mathbf{r}}{t}$ and  $\Cxrt{B}{\mathbf{r}}{t}$ denote the concentrations of type-$A$ and type-$B$ molecules in point $\mathbf{r}$ and at time $t$.  Considering a passive receiver, $\Ybxt{i}{t}$ is obtained as
\begin{IEEEeqnarray}{lll} \label{Eq:CR}
\Ybxt{i}{t} = \underset{\mathbf{r}\in\mathcal{V}^{\mathrm{rx}}}{\int} \Cxrt{i}{\mathbf{r}}{t} \mathrm{d}\mathbf{r},\quad i\in\{A,B\},
\end{IEEEeqnarray}
where $\mathcal{V}^{\mathrm{rx}}$ is the set of points within the receiver space.  Concentrations $\Cxrt{A}{\mathbf{r}}{t}$ and  $\Cxrt{B}{\mathbf{r}}{t}$  can be found using the following reaction-diffusion equations 
\cite{Nariman_Acid,ReactionDiffSim}
\begin{IEEEeqnarray}{lll} \label{Eq:Reaction_Diff}
\frac{\partial \Cxrt{A}{\mathbf{r}}{t}}{\partial t} = 
&D_A \nabla^2 \Cxrt{A}{\mathbf{r}}{t} \nonumber \\
&  - \kappa_f \Cxrt{A}{\mathbf{r}}{t}\Cxrt{B}{\mathbf{r}}{t} + \kappa_b+ \Gxrt{A}{\mathbf{r}}{t}\quad \IEEEyesnumber\IEEEyessubnumber \\
\frac{\partial \Cxrt{B}{\mathbf{r}}{t}}{\partial t} = 
&D_B \nabla^2 \Cxrt{B}{\mathbf{r}}{t} \nonumber \\
& - \kappa_f \Cxrt{A}{\mathbf{r}}{t}\Cxrt{B}{\mathbf{r}}{t} + \kappa_b + \Gxrt{B}{\mathbf{r}}{t}, \quad\,\,\, \IEEEyessubnumber 
\end{IEEEeqnarray}
where $\nabla^2=\frac{\partial^2}{\partial r_1^2}+\cdots+\frac{\partial^2}{\partial r_n^2},\,\,n\in\{1,2,3\}$, is the Laplace operator. Moreover,  $\Gxrt{i}{\mathbf{r}}{t}= \sum_{t_i\in\mathcal{T}_i} \Nxtx{i}  \delta_{\mathrm{d}}(\mathbf{r},t - t_i),\,\,i\in\{A,B\}$, represents the concentration of type-$i$ molecules that are released by the transmitter into the channel where $\delta_{\mathrm{d}}(\mathbf{r},t)=\delta_{\mathrm{d}}(r_1)\cdots\delta_{\mathrm{d}}(r_n)\delta_{\mathrm{d}}(t),\,\,n\in\{1,2,3\}$, and $\delta_{\mathrm{d}}(\cdot)$ is the Dirac delta function. In addition, without loss of generality, we assume  initial condition $\underset{\epsilon\to 0}{\lim}\,\,\Cxrt{i}{\mathbf{r}}{-\epsilon}=\Cinxr{i}{\mathbf{r}}$ where $\Cinxr{i}{\mathbf{r}}$ is the spatial concentration of type-$i$ molecules before transmission starts at $t=0$.

\begin{remk}\label{Remk:Equilib}
Suppose that the \ac{MC} environment is in steady state equilibrium before transmission starts at $t=0$. In this case, the concentrations of type-$A$ and type-$B$ molecules are temporally and spatially uniform, i.e., $\Cinxr{i}{\mathbf{r}}\triangleq C_i^{\mathrm{eq}}$. Therefore, substituting  $\frac{\partial \Cxrt{i}{\mathbf{r}}{t}}{\partial t} = 
D_i \nabla^2 \Cxrt{i}{\mathbf{r}}{t} =0$ into (\ref{Eq:Reaction_Diff}), we obtain that the following relation holds for the equilibrium concentration \cite[Chapter~14]{Tro2015Chemistry}
\begin{IEEEeqnarray}{lll} \label{Eq:InitialConcentrationKw}
C_A^{\mathrm{eq}}C_B^{\mathrm{eq}} = \frac{\kappa_b}{\kappa_f}\triangleq \kappa_w.
\end{IEEEeqnarray}
For instance, assuming that type-$A$ and type-$B$ molecules represent acid and base and the fluid is water, we have $\kappa_f=1.4\times 10^{11} \,\,1/(\text{Ms})$ and $\kappa_b=1.4\times 10^{-3}\,\,\text{M}/\text{s}$ at $25^{\circ}$C which leads to $\kappa_w=10^{-14}$~M$^2$ where M, called molar, is a unit of concentration and represents the number of moles per every liter of solution in a three-dimensional environment\footnote{The concentration $C$ in molecule$\cdot$m$^{-3}$ is related to the molar concentration $C^{\mathrm{mol}}$ in mol$\cdot$L$^{-1}$ via $C= 10^3 n_{\mathrm{av}}C^{\mathrm{mol}}$ where $n_{\mathrm{av}}=6.02\times 10^{23}$ is the Avogadro constant \cite{Tro2015Chemistry}.} \cite[Chapter~14]{Tro2015Chemistry}.  Therefore, when the concentrations of acid and base are identical, i.e., $C_A^{\mathrm{eq}}=C_B^{\mathrm{eq}}=10^{-7}$ M, the pH, defined by $\text{pH} = -\log_{10}(C_A^{\mathrm{eq}})$, is $7$.  As can be seen from (\ref{Eq:InitialConcentrationKw}), regardless of whether or not molecules are released by the transmitter into the channel, some concentrations of type-$A$ and type-$B$ molecules are present in the environment which can be interpreted as environmental noise molecules. However, unlike non-reactive \ac{MC} systems where the concentrations of signaling and noise molecules in the environment are typically assumed to be independent, for reactive \ac{MC} systems, the concentrations of noise and signaling molecules are not independent and are jointly described by (\ref{Eq:Reaction_Diff}).
\end{remk}

As stated earlier, the general reaction-diffusion equations in (\ref{Eq:Reaction_Diff}) have not yet been solved in closed form. The main difficulty in doing so arises from  the coupling of the two equations and the non-linear term $ \kappa_f \Cxrt{A}{\mathbf{r}}{t}\Cxrt{B}{\mathbf{r}}{t}$. Note that even if we assume that one of the variables, e.g. $C_B(\mathbf{r},t)$, is fixed, it is still challenging to solve (\ref{Eq:Reaction_Diff}a) in terms of $C_A(\mathbf{r},t)$.  Therefore, in the following, we derive a numerical method for solving (\ref{Eq:Reaction_Diff}) in a computationally efficient manner. This is achieved by fully exploiting the characteristics of the considered system model. 

\subsection{Derivation of the CR}\label{Sec:CR}

Let us assume that time is divided into a series of small intervals of length $\Delta t$. The main idea behind the proposed approach for computing the \ac{CR} is to find the concentrations at the end of each time interval given the concentrations at the beginning of the time interval  while exploiting the condition $\Delta t\to 0$. In particular, from the reaction-diffusion equations in (\ref{Eq:Reaction_Diff}), we have
\begin{IEEEeqnarray}{lll} \label{Eq:Reaction_Diff_integral}
\Cxrt{i}{\mathbf{r}}{t+\Delta t} = &\,
\Cxrt{i}{\mathbf{r}}{t} + \int_{\tilde{t}=t}^{t+\Delta t} G_i(\mathbf{r},t+\Delta t)\mathrm{d}\tilde{t} \nonumber \\
& +\underset{\Delta C_i^{\mathrm{df}}(\mathbf{r},t+\Delta t)}{\underbrace{\int_{\tilde{t}=t}^{t+\Delta t}  D_i \nabla^2 C_i(\mathbf{r},\tilde{t}) \mathrm{d}\tilde{t}}}  \nonumber \\
&+  \underset{\Delta C_i^{\mathrm{rc}}(\mathbf{r},t+\Delta t)}{\underbrace{\int_{\tilde{t}=t}^{t+\Delta t}   \left(\kappa_b- \kappa_f C_A(\mathbf{r},\tilde{t})C_B(\mathbf{r},\tilde{t}) \right) \mathrm{d}\tilde{t}}},\,\,\,
\quad
\end{IEEEeqnarray}
where  $\Delta C_i^{\mathrm{df}}(\mathbf{r},t+\Delta t)$ and $\Delta C_i^{\mathrm{rc}}(\mathbf{r},t+\Delta t)$ are concentration changes from time $t$ to time $t+\Delta t$ due to  diffusion and (forward and backward) reaction, respectively. Note that, in general, the diffusion and reaction processes are coupled which makes the derivation of a closed-form expression for $\Delta C_i^{\mathrm{x}}(\mathbf{r},t+\Delta t),\,\,\mathrm{x}\in\{\mathrm{df},\mathrm{rc}\},$ intractable. Nevertheless, the following lemma specifies $\Delta C_i^{\mathrm{x}}(\mathbf{r},t+\Delta t)$ for $\Delta t\to 0$. First, let us formally introduce the following two sets of assumptions which are needed for a rigorous statement of Lemma~\ref{Lem:Numerical}: A1) We assume $C_i(\mathbf{r},t)\neq 0$ and $\nabla^2 C_i(\mathbf{r},t)\neq 0$ hold for $\forall \mathbf{r},t$. A2) We assume $\mathcal{T}_A\cap\mathcal{T}_B$ is empty and that $\mathcal{T}_i\subset\mathcal{T}$ where  $\mathcal{T}=\{t|t=m\Delta t - \varepsilon,\,\,m\in\mathbb{N}\}$, $\mathbb{N}$ is the set of natural numbers, and $\varepsilon$ is an arbitrary small positive real number satisfying $\varepsilon\ll\Delta t$. Note that assumption A1 is mild as it generally holds after the first release of molecules by the transmitter for any point $\mathbf{r}$ in the MC environment. Similarly, assumption A2 is not limiting since $\Delta t$ can be chosen to be sufficiently small  to accommodate any desired release time pattern of molecules with high accuracy while satisfying A2.

\begin{lem}\label{Lem:Numerical} Under assumptions A1 and A2, the following asymptotic result holds for  $\Delta C_i^{\mathrm{x}}(\mathbf{r},t+\Delta t)$:
\begin{IEEEeqnarray}{lll}\label{Eq:Separation}  
\underset{\Delta t\to 0}{\lim}\,\,\Delta C_i^{\mathrm{x}}(\mathbf{r},t+\Delta t) = C_i^{\mathrm{x}}(\mathbf{r},t+\Delta t) - C_i(\mathbf{r},t), \mathrm{x}\in\{\mathrm{df},\mathrm{rc}\},\quad
\end{IEEEeqnarray}
where $C_i^{\mathrm{df}}(\mathbf{r},t)$ and $C_i^{\mathrm{rc}}(\mathbf{r},t)$ are the type-$i$ molecule concentration assuming that in interval $[t,t+\Delta t]$, \textit{only} diffusion and \textit{only} reactions occur, respectively,  and the other phenomenon is absent. 
\end{lem}
\begin{IEEEproof}
The proof is given in Appendix~\ref{App:Lem_Numerical}.
\end{IEEEproof}

The main result in the above lemma is that assuming $\Delta t\to 0$, we can decouple the impact of diffusion and reaction and compute the concentration change due to each phenomenon separately. More specifically, the above lemma indicates that the effect of the coupling of the involved phenomena on the concentration change has a negligible impact compared to the effect of the individual phenomena as  $\Delta t\to 0$. The decoupling of reaction and diffusion equations in (\ref{Eq:Separation}) can be seen as a form of general splitting methods used in numerical mathematics \cite{besse2002order}, \cite{blanes2008splitting}, \cite{hundsdorfer2013numerical}. Based on this result, the concentration update rule for $\Cxrt{i}{\mathbf{r}}{t+\Delta t}$ for the considered \ac{MC} system is given in the following proposition.

\begin{prop}\label{Prop:Numerical}
Under assumptions A1 and A2, $\Cxrt{i}{\mathbf{r}}{t+\Delta t}$ is obtained as
\begin{IEEEeqnarray}{lll} \label{Eq:UpdateRule}
	\Cxrt{i}{\mathbf{r}}{t+\Delta t} \nonumber \\
	= \bar{G}_i(\mathbf{r})  + C_i^{\mathrm{df}}(\mathbf{r},t+\Delta t) + C_i^{\mathrm{rc}}(\mathbf{r},t+\Delta t)- \Cxrt{i}{\mathbf{r}}{t},
\end{IEEEeqnarray}
where $\bar{G}_i(\mathbf{r})= \sum_{t_i\in\mathcal{T}_i} \Nxtx{i}  \delta_{\mathrm{d}}(\mathbf{r})\delta_{\mathrm{k}}(t+\Delta t-\varepsilon-t_i)$ where $\delta_{\mathrm{k}}(\cdot)$ is the Kronecker delta function. Moreover, for the \ac{MC} system under consideration, $C_i^{\mathrm{df}}(\mathbf{r},t+\Delta t)$ and $C_i^{\mathrm{rc}}(\mathbf{r},t+\Delta t)$ are given by (\ref{Eq:DiffRctSol}) on the top of the next page  
where $c_1(\mathbf{r},t)=C_A(\mathbf{r},t)-C_B(\mathbf{r},t)$, $c_2(\mathbf{r},t)=\sqrt{\kappa_f^2c_1^2(\mathbf{r},t)+4\kappa_f\kappa_b}$,  $c_3(\mathbf{r},t)=C_A(\mathbf{r},t)+C_B(\mathbf{r},t)$, and $c_4(\mathbf{r},t)=\frac{c_2(\mathbf{r},t)-\kappa_fc_3(\mathbf{r},t)}{c_2(\mathbf{r},t)+\kappa_fc_3(\mathbf{r},t)}$. 
\begin{figure*}
\begin{IEEEeqnarray}{rll}\label{Eq:DiffRctSol}
	C_i^{\mathrm{df}}(\mathbf{r},t+\Delta t)  &=    
	\frac{1}{(4\pi D_i \Delta t)^{\frac{n}{2}}} 
	\iiint_{\tilde{\mathbf{r}}}
	C_i(\tilde{\mathbf{r}},t) \exp\left(-\frac{\|\mathbf{r}-\tilde{\mathbf{r}}\|^2}{4D_i \Delta t}\right)\mathrm{d}\tilde{\mathbf{r}} \IEEEyesnumber\IEEEyessubnumber  \\
	C_A^{\mathrm{rc}}(\mathbf{r},t+\Delta t) &= 
	\frac{c_2(\mathbf{r},t)+\kappa_fc_1(\mathbf{r},t)-\big(c_2(\mathbf{r},t)-\kappa_fc_1(\mathbf{r},t)\big)
		c_4(\mathbf{r},t)\exp\big(-c_2(\mathbf{r},t)\Delta t\big)} 
	{2\kappa_f\left[1+c_4(\mathbf{r},t)\exp\big(-c_2(\mathbf{r},t)\Delta t\big)\right]} \IEEEyessubnumber  \\
	C_B^{\mathrm{rc}}(\mathbf{r},t+\Delta t) &= 
	\frac{c_2(\mathbf{r},t)-\kappa_fc_1(\mathbf{r},t)-\big(c_2(\mathbf{r},t)+\kappa_fc_1(\mathbf{r},t)\big)
		c_4(\mathbf{r},t)\exp\big(-c_2(\mathbf{r},t)\Delta t\big)} 
	{2\kappa_f\left[1+c_4(\mathbf{r},t)\exp\big(-c_2(\mathbf{r},t)\Delta t\big)\right]}. \IEEEyessubnumber
\end{IEEEeqnarray}
\hrulefill
\end{figure*}
\end{prop}
\begin{IEEEproof}
The proof is given in Appendix~\ref{App:Prop_Numerical}.
\end{IEEEproof}

Algorithm~\ref{Alg:CR} summarizes the simulation steps for \ac{CR} computation using Proposition~\ref{Prop:Numerical} where $T^{\max}$ is the maximum simulation time.

\begin{algorithm}[t]
\caption{Computation of CR}
 \begin{algorithmic}[1]\label{Alg:CR}
 \STATE \textbf{initialize:} $t=0$, $\Delta t$, $T^{\max}$, $\mathcal{T}_i$, and $C_{i}(\mathbf{r},0)$.   
      \WHILE{$t\leq T^{\max}$}
              \STATE Update $t$ with  $t+\Delta t$.
              \STATE Compute $C_i^{\mathrm{x}}(\mathbf{r},t)$ based on (\ref{Eq:DiffRctSol}).
        \STATE Update $C_{i}(\mathbf{r},t)$ according to (\ref{Eq:UpdateRule}).
      \ENDWHILE
      \STATE Return $\bar{y}_i(t)$ from (\ref{Eq:CR}) as the CR.
  \end{algorithmic}
\end{algorithm}

\begin{remk}\label{Remk:FirstOrderApprox}
We note that as $\Delta t\to 0$, one can also use the following first order approximations for $C_i^{\mathrm{df}}(\mathbf{r},t+\Delta t)$ and $C_i^{\mathrm{rc}}(\mathbf{r},t+\Delta t)$ respectively:
\begin{IEEEeqnarray}{rll}\label{Eq:FirstOrderApprox}
C_i^{\mathrm{df}}(\mathbf{r},t+\Delta t)  &= C_i(\mathbf{r},t) +  D_i \nabla^2 C_i(\mathbf{r},t) \Delta t, \IEEEyesnumber\IEEEyessubnumber \\
C^{\mathrm{rc}}_{i}(\mathbf{r},t+\Delta t) &= C_i(\mathbf{r},t) + \big(\kappa_b- \kappa_f C_A(\mathbf{r},t)C_B(\mathbf{r},t)\big)\Delta t. \IEEEyessubnumber
\end{IEEEeqnarray}
 However, for these approximations to be valid, $\Delta t$ has to be chosen much smaller than when the exact expressions in Proposition~\ref{Prop:Numerical} are adopted. Moreover, small values of $\Delta t$ lead to stability issues for computation of $\nabla^2 C_i(\mathbf{r},t)$, see Section~\ref{Sec:Complexity} for a detailed discussion.  
\end{remk}

\subsection{Simplifications for Special Cases}

In the following, we further simplify the diffusion term $C_i^{\mathrm{df}}(\mathbf{r},t+\Delta t)$ and reaction term $C_i^{\mathrm{rc}}(\mathbf{r},t+\Delta t)$ for the considered system model.

\textit{Diffusion:} The computational complexity of Algorithm~\ref{Alg:CR} is largely determined by (\ref{Eq:DiffRctSol}a) since for each update, an $n$-dimensional integral has to be evaluated for each point $\mathbf{r}\in\mathbb{R}^n$ where $\mathbb{R}$ is the set of real numbers.  Nevertheless, for the commonly adopted assumption of a point-source transmitter with impulsive release \cite{Adam_Enzyme,Nariman_Acid,Reza_Reaction},  the computation of $C_i^{\mathrm{df}}(\mathbf{r},t)$  can be significantly simplified using the following corollary.
 
\begin{corol}\label{Corol:OneD}
For the \ac{MC} system under consideration, the concentrations of the molecules are symmetric with respect to origin and hence are only functions of variable $r\triangleq \|\mathbf{r}\|$. In this case,  $C_i^{\mathrm{df}}(\mathbf{r},t+\Delta t) $ can be simplified as
\begin{IEEEeqnarray}{rll} \label{Eq:Concen_OneD}
C_i^{\mathrm{df}}(\mathbf{r},t+\Delta t)  \,
&=  \frac{1}{(4\pi D_i \Delta t)^{n/2}} 
\int_{\tilde{r}=0}^{\infty} C_i(\tilde{r},t) 
W_i(r,\tilde{r}) \mathrm{d}\tilde{r},\quad
\end{IEEEeqnarray}
where $W_i(r,\tilde{r})$ is given by
\begin{IEEEeqnarray}{lll} \label{Eq:WeightFunc}
W_i(r,\tilde{r}) = &2 
\exp\left(-\frac{\tilde{r}^2+r^2}{4D_i \Delta t}\right) \nonumber \\
&\times
\begin{cases}
\cosh\left(\frac{r\tilde{r}}{2D_i\Delta t}\right), &\mathrm{if} \,\,n=1\\
\pi\tilde{r} I_0\left(\frac{r\tilde{r}}{2D_i \Delta t}\right), &\mathrm{if} \,\,n=2\\
\frac{4\pi D_i\tilde{r} \Delta t}{r}\sinh\left(\frac{r\tilde{r}}{2D_i\Delta t}\right), &\mathrm{if} \,\,n=3
\end{cases}.
\end{IEEEeqnarray}
Here, $I_0(\cdot)$ is the zeroth order modified Bessel function of the first kind, $\sinh(\cdot)$ is the sine hyperbolic function, and   $\cosh(\cdot)$ is the cosine hyperbolic function.
\end{corol}
\begin{IEEEproof}
The proof is given in Appendix~\ref{App:Corol_OneD}.
\end{IEEEproof}

Note that the $n$-dimensional integral in (\ref{Eq:DiffRctSol}a) is simplified to a one-dimensional integral in (\ref{Eq:Concen_OneD}) which has to be evaluated for a one-dimensional space variable $r\in\mathbb{R}^+$, where $\mathbb{R}^+$ denotes the set of non-negative real numbers. In addition, the term $W_i(r,\tilde{r})$ in the integral  does not depend on the molecule concentrations. Hence, we can evaluate it offline and use it for online concentration updates. In other words, the integral in (\ref{Eq:Concen_OneD}) can be approximated by summation and multiplication operations.  

\textit{Reaction:} In certain cases, only the forward reaction (degradation) or the backward reaction (production) may be present. To this end, one must evaluate the limits of (\ref{Eq:DiffRctSol}b) and (\ref{Eq:DiffRctSol}c) for either $\kappa_f\to 0$ or $\kappa_b \to 0$. In particular, (\ref{Eq:DiffRctSol}b) and (\ref{Eq:DiffRctSol}c) have indeterminate forms at $\kappa_f=0$. Moreover, when $C_A(\mathbf{r},t)=C_B(\mathbf{r},t)$ holds, (\ref{Eq:DiffRctSol}b) and (\ref{Eq:DiffRctSol}c) have indeterminate forms at $\kappa_b=0$, too. The following corollary provides the results for these cases.

\begin{corol}\label{Corol:FrBcSep}
For the considered \ac{MC} system, the molecule concentrations when only the forward reaction is present, denoted by $C^{\mathrm{fr}}_{i}(\mathbf{r},t+\Delta t)$, and when only the backward reaction is present, denoted by $C^{\mathrm{br}}_{i}(\mathbf{r},t+\Delta t)$, are given by
\begin{IEEEeqnarray}{rll}
C^{\mathrm{fr}}_{A}(\mathbf{r},t+\Delta t) & =\underset{\kappa_b\to 0}{\lim} \,\, C^{\mathrm{rc}}_{A}(\mathbf{r},t+\Delta t) \nonumber \\
&= \begin{cases}
\frac{c_1(\mathbf{r},t)}{1-c_5(\mathbf{r},t)\mathrm{exp}\left(-\kappa_f c_1(\mathbf{r},t)\Delta t\right)}, & \mathrm{if}\,\, c_1(\mathbf{r},t)\neq 0 \\ 
\frac{1}{\kappa_f t +\frac{1}{C_A(\mathbf{r},t)}}, &\mathrm{if}\,\,c_1(\mathbf{r},t) = 0
\end{cases} \IEEEyesnumber \IEEEyessubnumber \label{Eq:ForRSol_A} \\
C^{\mathrm{fr}}_{B}(\mathbf{r},t+\Delta t) & = \underset{\kappa_b\to 0}{\lim} \,\, C^{\mathrm{rc}}_{B}(\mathbf{r},t+\Delta t) \nonumber \\
&= \begin{cases}
\frac{c_1(\mathbf{r},t)c_5(\mathbf{r},t)\mathrm{exp}\left(-\kappa_f c_1(\mathbf{r},t)\Delta t\right)}{1-c_5(\mathbf{r},t)\mathrm{exp}\left(-\kappa_f c_1(\mathbf{r},t)\Delta t\right)}, & \mathrm{if}\,\, c_1(\mathbf{r},t)\neq 0 \\ 
\frac{1}{\kappa_f t +\frac{1}{C_B(\mathbf{r},t)}}, &\mathrm{if}\,\,c_1(\mathbf{r},t) = 0
\end{cases} \qquad \IEEEyessubnumber \label{Eq:ForRSol_B} \\
C_i^{\mathrm{br}}(\mathbf{r},t+\Delta t)  & = \underset{\kappa_f\to 0}{\lim} \,\, C^{\mathrm{rc}}_{i}(\mathbf{r},t+\Delta t) \nonumber \\
&= C_i(\mathbf{r},t) + \kappa_b \Delta t,\quad i\in\{A,B\}, \label{Eq:BackRSol}
\end{IEEEeqnarray}
where $c_5(\mathbf{r},t)=\frac{C_B(\mathbf{r},t)}{C_A(\mathbf{r},t)}$. 
\end{corol}
\begin{IEEEproof}
Due to the indeterminate form of $C^{\mathrm{rc}}_{i}(\mathbf{r},t+\Delta t)$ at $\kappa_b=0$, we use L'Hopital's rule \cite{Taylor1952hospital} to compute   $C^{\mathrm{fr}}_{i}(\mathbf{r},t+\Delta t)=\underset{\kappa_b\to 0}{\lim} \,\, C^{\mathrm{rc}}_{i}(\mathbf{r},t+\Delta t)$ where $C^{\mathrm{rc}}_{i}(\mathbf{r},t+\Delta t)$ is given by (\ref{Eq:DiffRctSol}b) and (\ref{Eq:DiffRctSol}c). This leads to  $C^{\mathrm{fr}}_{i}(\mathbf{r},t+\Delta t)$ given in (\ref{Eq:ForRSol_A}) and (\ref{Eq:ForRSol_B}) for the case when $c_1(\mathbf{r},t)\neq 0$. Note that these expressions for $C^{\mathrm{fr}}_{i}(\mathbf{r},t+\Delta t)$ are indeterminate at $c_1(\mathbf{r},t)= 0$. In a similar manner, we apply L'Hopital's rule to $C^{\mathrm{fr}}_{i}(\mathbf{r},t+\Delta t)$ given in (\ref{Eq:ForRSol_A}) and (\ref{Eq:ForRSol_B}) for $c_1(\mathbf{r},t)\neq 0$ in order to find the corresponding expressions when $c_1(\mathbf{r},t)= 0$. For the case, when $\kappa_f=0$, it is easier to solve the original PDEs in (\ref{Eq:Reaction_Diff}) which simplifies to $\frac{\partial C_i^{\mathrm{br}}(\mathbf{r},t)}{\partial t}=\kappa_b$ and has the immediate solution given in (\ref{Eq:BackRSol}). This completes the proof.
\end{IEEEproof}

\subsection{Discussion on Complexity}\label{Sec:Complexity}

In the following, we compare the computational complexity of Algorithm~\ref{Alg:CR} with that of the conventional numerical methods such as \ac{FDM}. Most numerical methods in the literature rely on discretization of space and time to solve the reaction-diffusion equations in (\ref{Eq:Reaction_Diff}) \cite{PDE_numerical,Nariman_Acid}. In particular, for \ac{FDM}, time  and space are discretized into small intervals to approximate the differential operators in (\ref{Eq:Reaction_Diff}). Let $N_t$ and $N_r$ denote the number of time steps and the number of space points per dimension used for discretization of time and space, respectively. The advantage of numerical methods such as \ac{FDM} is their universality as they can be applied to general \acp{PDE}. However, for the approximation of the differential operators to be accurate, the adopted step sizes should be very small, i.e., $N_t$ and $N_r$ should be very large \cite{PDE_numerical}. However, increasing $N_t$ and $N_r$ too much results in numerical instability for computation of the differential operators, e.g., $\nabla^2 C_i(\mathbf{r},t)$. In fact, it is widely documented in the numerical mathematics literature that the overall accuracy and stability of \ac{FDM} methods heavily depend on proper selection of the step sizes $N_t$ and $N_r$ \cite{PDE_numerical,Nariman_Acid}. Moreover, the appropriate values of the step sizes are application-specific and there is no universal guideline that can be applied to all \acp{PDE} and system parameters.  On the contrary, the proposed approach in Algorithm~\ref{Alg:CR} is a hybrid method where time is discretized; however, the problem is solved analytically in (\ref{Eq:DiffRctSol}) and (\ref{Eq:Concen_OneD}) with respect to the space variables. Since no differential operator is used in these update rules, Algorithm~\ref{Alg:CR} does not face the stability issues of \ac{FDM} methods. Moreover, compared to \ac{FDM} methods,  larger step sizes can be adopted in the proposed algorithm for the same overall accuracy. This is due to the fact that unlike \ac{FDM} methods, which approximate the differential operators in (\ref{Eq:Reaction_Diff}), we solve them exactly for the diffusion-only and reaction-only scenarios and errors occur only due to ignoring the coupling of reaction and diffusion in (\ref{Eq:Reaction_Diff}).  This makes the proposed algorithm much faster than pure numerical methods such as FDM and avoids the stability issues that arise in the FDM methods.

\begin{remk}
In this paper, we do not quantify the relative complexities of FDM methods and the proposed algorithm since this comparison depends on the specific adopted implementation of the FDM method and the values of the system parameters. However, in this remark, we discuss how the complexity of these methods scale in terms of $N_t$ and $N_r$. In particular, since the molecule concentrations have to be updated in each time step and for each space point, the computational complexity of FDM methods has   order $\mathcal{O}(N_tN_r^n)$ where $\mathcal{O}(\cdot)$ is the standard big-O notation. Moreover, if (\ref{Eq:Reaction_Diff}) is rewritten in polar and spherical coordinates when $n=2$ and $3$, respectively, the computational complexity can be reduced to $\mathcal{O}(N_tN_r)$ by exploiting symmetry with respect to $\|\mathbf{r}\|$. The complexity order of Algorithm~\ref{Alg:CR} is given by $\mathcal{O}(N_tN_r^{2n})$. However, by exploiting the symmetry and using the update rule for $C_i^{\mathrm{df}}(\mathbf{r},t)$ given in (\ref{Eq:Concen_OneD}), the complexity of Algorithm~\ref{Alg:CR}  is reduced to $\mathcal{O}(N_tN_r^2),\,\,n=1,2,3$. Furthermore, if one employs the first order approximations in (\ref{Eq:FirstOrderApprox}) as concentration update rules in Algorithm~\ref{Alg:CR},  complexity is further reduced to $\mathcal{O}(N_tN_r)$. However, the implementation of the concentration update rules in (\ref{Eq:FirstOrderApprox}) suffers the same stability issues as the FDM methods. In fact, the main advantage of the proposed algorithm using the updates rules in (\ref{Eq:DiffRctSol}) and (\ref{Eq:Concen_OneD}) is that it does not face the stability issues  of FDM methods. Moreover, the required $N_t$ and $N_r$ are much smaller than those needed for FDM methods. Although for given $N_t$ and $N_r$, the complexity order of the proposed algorithm is higher than that of FDM methods. As a remedy to this issue, one may compute the integration in (\ref{Eq:DiffRctSol}a) over $\tilde{\mathbf{r}}\in\mathcal{R}(\mathbf{r})$, instead of over the entire space where $\mathcal{R}(\mathbf{r})=\big\{\tilde{\mathbf{r}}\big|\|\tilde{\mathbf{r}}-\mathbf{r}\|\leq R\big\}$ and $R$ is a parameter referred to as integration radius. Thereby, the complexity order of the proposed algorithm reduces to $\mathcal{O}(N_tN_r^n)$ and $\mathcal{O}(N_tN_r)$ for concentrations that are asymmetric and symmetric with respect to the origin, respectively, i.e., the same complexity order as for FDM methods. 
\end{remk}

\begin{table}
	\label{Table:Parameter}
	\caption{Default Values of the System Parameters \cite{Adam_Enzyme,Nariman_Acid}. \vspace{-0.2cm}} 
	\begin{center}
		\scalebox{0.65}
		{
			\begin{tabular}{|| c | c | c ||}
				\hline 
				Variable & Definition & Value \\ \hline \hline
				$N^{\mathrm{tx}}_A,N^{\mathrm{tx}}_B$ & Number of released molecules  & $5\times 10^3$ molecules \\ \hline      
				$d$ &  Distance between  transmitter and  receiver  & $250$ nm\\ \hline 
				$D_A,D_B$ &  Diffusion coefficient& $ 10^{-10}$ $\text{m}^2\cdot\text{s}^{-1}$\\ \hline          
				$\kappa_f$ &  Forward reaction rate  & $10^{-17}$  molecule$^{-1}\cdot$m$^3\cdot$s$^{-1}$\\   \hline
				$\kappa_b$ &  Backward reaction rate  & $10^{25}$  molecule$\cdot$m$^{-3}\cdot$s$^{-1}$\\   \hline
				$r$ & Receiver radius   & $50$ nm \\    \hline 
				$T^{\mathrm{symb}}$ &  Symbol duration  & $200$  $\mu$s\\   \hline
			\end{tabular}
		}
	\end{center}\vspace{-0.6cm}
\end{table}

\section{Simulation Results}\label{Sec:SimResult}

In this section, we first verify the accuracy of the proposed \ac{CR} computation algorithm and the Poisson statistics assumed in (\ref{Eq:Poisson_ISI}) using the particle-based simulator described in \iftoggle{arXiv}{Appendix~F.}{\cite[Appendix~F]{Reactive_arXiv}.} Next, the effect of some important system parameters on the \ac{CR} is studied, and finally, the performance of the proposed modulation schemes and detectors is evaluated in terms of the \ac{BER}. For clarity of presentation, we focus on a three-dimensional environment and unless stated otherwise, the default values for the system parameters given in Table~I are used.  The time step size is  $\Delta t=1\,\mu$s for Algorithm~\ref{Alg:CR} and the particle-based simulator, respectively. Moreover, the simulation results shown in Figs.~\ref{Fig:CR_eq} and \ref{Fig:Poisson_PDF} are obtained by  running the particle-based simulator  $10^4$ times.

\subsection{Verification of the Proposed CR Computation Algorithm and the Poisson Statistics}

\begin{figure} 
  \centering  
\resizebox{0.9\linewidth}{!}{\psfragfig{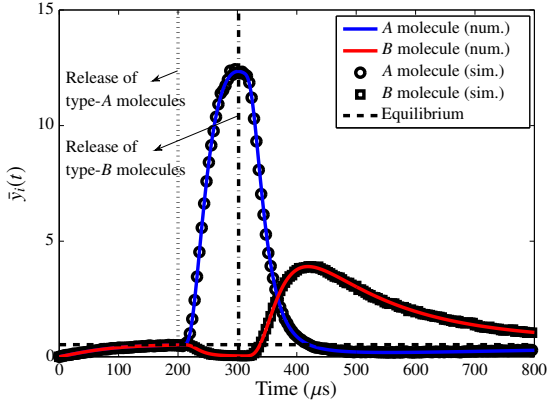}} \vspace{-0.3cm}
\caption{CRs $\bar{y}_A(t)$ and $\bar{y}_B(t)$. The vertical dotted and dashed-dotted lines denote the release times of type-$A$ and type-$B$ molecules, respectively.  }
\label{Fig:CR_eq}
\end{figure}

In Fig.~\ref{Fig:CR_eq}, we plot the \textit{expected} number of molecules observed at the receiver, i.e., CR, where $\mathcal{T}_A=\{200\}$ and $\mathcal{T}_B=\{300\}$ $\mu$s and for initial condition $C_A^{\mathrm{init}}(\mathbf{r})=C_B^{\mathrm{init}}(\mathbf{r})=0$. We show numerical results obtained with Algorithm~\ref{Alg:CR}  and simulation results generated with the particle-based simulator introduced in Appendix~\ref{Sec:SimParticle}. From Fig.~\ref{Fig:CR_eq}, we observe a perfect agreement between numerical and simulation results which validates the accuracy of the proposed numerical algorithm. In Fig.~\ref{Fig:CR_eq}, we also plot the equilibrium concentration which can be computed based on Remark~\ref{Remk:Equilib} as $\bar{y}_i^{\mathrm{eq}}=C_i^{\mathrm{eq}}V^{\mathrm{rx}}=0.5236$ where $C_A^{\mathrm{eq}}=C_B^{\mathrm{eq}}=\sqrt{\frac{\kappa_b}{\kappa_f}}$ and $V^{\mathrm{rx}}$ is the receiver volume. Note that we assume no molecules are present in the environment at $t=0$ and the transmitter also releases no molecule before $t=200\mu$s. Nevertheless, $\bar{y}_i(t)$ increases in interval $t\in[0,200\mu\text{s}]$ due to the backward reaction (production) in (\ref{Eq:Reaction}) and finally converges to $\bar{y}_i(t)=\bar{y}_i^{\mathrm{eq}}$ since the backward reaction and the forward reaction (degradation) in (\ref{Eq:Reaction}) reach an equilibrium, cf. Remark~\ref{Remk:Equilib}. 

\begin{figure} 
		\centering  
		\resizebox{0.9\linewidth}{!}{\psfragfig{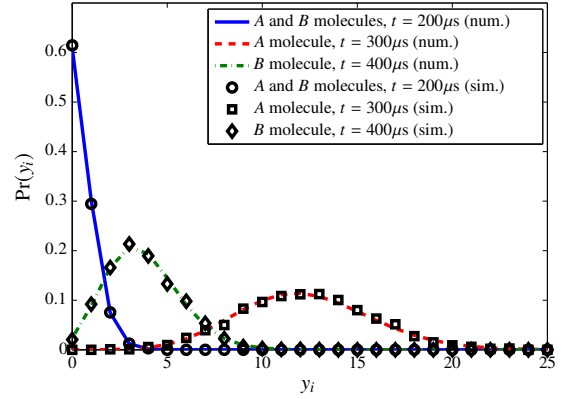}} \vspace{-0.3cm}
		\caption{PMFs of observations $y_A$ and $y_B$ at three sample times $t=200,300,400$ $\mu$s in Fig.~\ref{Fig:CR_eq}.}
		\label{Fig:Poisson_PDF}
\end{figure}

Next, we evaluate the accuracy of the Poisson statistical model for $y_i(t)$ introduced in (\ref{Eq:Poisson_ISI}), where the mean is obtained using Algorithm~\ref{Alg:CR}, by comparing it with the histogram of $y_i(t)$, which is obtained via particle-based simulation, i.e., the ``exact distribution". In particular, in Fig.~\ref{Fig:Poisson_PDF}, we show the \ac{PMF} of $y_i(t)$ for a few different time instances, namely $t=200,300$ $\mu$s for type-$A$ molecules and $t=200,400$ $\mu$s for type-$B$ molecules. The same parameters as for Fig.~\ref{Fig:CR_eq} are adopted for Fig.~\ref{Fig:Poisson_PDF}.  Note that the \ac{PMF}s of $y_A(t)$ and $y_B(t)$ are identical at $t=200$ $\mu$s, i.e., before the release of molecules by the transmitter, due to the identical characteristics of the type-$A$ and type-$B$ molecules.
Moreover, we emphasize that for the considered time instances, the mean of $y_i(t)$ assumes a small value $\bar{y}_A(t=200\mu\text{s})=\bar{y}_B(t=200\mu\text{s})\approx 0.5$, a moderate value $\bar{y}_B(t=400\mu\text{s})\approx 4$, and a large value $\bar{y}_A(t=300\mu\text{s})\approx 12$, and is given by the \ac{CR} plotted in Fig.~\ref{Fig:CR_eq}. As can be seen from Fig.~\ref{Fig:Poisson_PDF}, the Poisson statistics with the mean obtained by Algorithm~\ref{Alg:CR} can accurately model the histograms obtained from particle-based simulation for the considered small, moderate, and large mean values. This validates the Poisson model for \ac{MC} systems with reactive signaling and is in line with the results reported in the chemistry and physics literature \cite{PoissonGardiner,CoxNatureCommun}. 

\subsection{Effect of System Parameters on CR}

In this subsection, we study the impact of the diffusion coefficient, forward reaction rate, and backward reaction rate on the CR.  

\begin{figure}
  \centering  
\resizebox{0.9\linewidth}{!}{\psfragfig{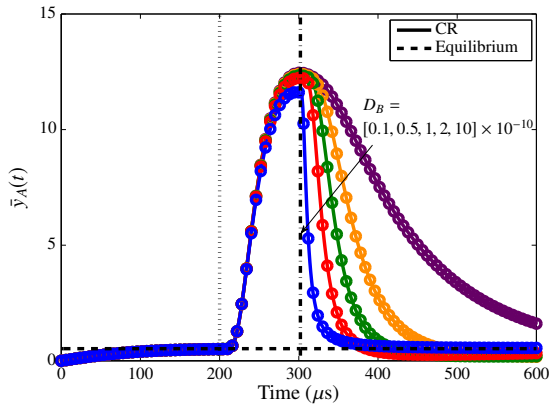}}  \vspace{-0.3cm}
\caption{CR $\bar{y}_A(t)$ for $D_A=10^{-10}$ m$^2\cdot$s$^{-1}$ and $D_B=[0.1, 0.5, 1, 2, 10]\times 10^{-10}$ m$^2\cdot$s$^{-1}$. The vertical dotted and dashed-dotted lines denote the release times of type-$A$ and type-$B$ molecules, respectively.  }
\label{Fig:CR_DB}
\end{figure}

In Fig.~\ref{Fig:CR_DB}, we plot the \ac{CR} for type-$A$ molecules, $\bar{y}_A(t)$,  versus time for $\mathcal{T}_A=\{200\}$ and $\mathcal{T}_B=\{300\}$ $\mu$s, and initial condition $C_A^{\mathrm{init}}(\mathbf{r})=C_B^{\mathrm{init}}(\mathbf{r})=0$. We consider a fixed $D_A=10^{-10}$ m$^2\cdot$s$^{-1}$ and different $D_B=[0.1, 0.5, 1, 2, 10]\times 10^{-10}$ m$^2\cdot$s$^{-1}$. As can be seen from this figure, increasing the diffusion coefficient of the type-$B$ molecules shortens the \ac{CR}. This is due to the fact that the cleaning type-$B$ molecules reach and cancel out the previously released type-$A$ molecules faster when $D_B$ is larger, i.e., the \ac{ISI} reduction is more effective. Moreover, as expected, the equilibrium level of \ac{CR} $\bar{y}_A^{\mathrm{eq}}$ is not affected by the value of $D_B$; however, a larger $D_B$ leads to a faster convergence to this equilibrium~level.

\begin{figure}
		\centering  
		\resizebox{0.9\linewidth}{!}{\psfragfig{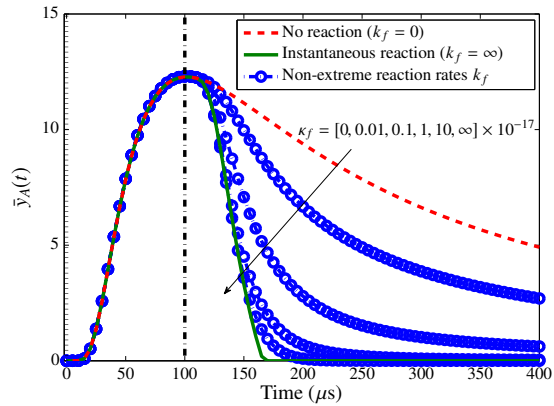}}  \vspace{-0.3cm}
		\caption{CR $\bar{y}_A(t)$ for $k_f=[0, 0.01, 0.1, 1, 10, \infty]\times 10^{-17}$ molecule$^{-1}\cdot$m$^3\cdot$s$^{-1}$ and $k_b=0$.  The vertical dotted and dashed-dotted lines denote the release times of type-$A$ and type-$B$ molecules, respectively.  }
		\label{Fig:CR_kf}
\end{figure}

Next, we study the effect of forward reaction rate constant $\kappa_f$ on the \ac{CR}. We adopt $\kappa_b=0$ and $\kappa_f\in\{0,0.01,0.1,1,10,\infty\}\times 10^{-17}$ molecule$^{-1}\cdot$m$^3\cdot$s$^{-1}$ where the extreme values of $\kappa_f=0,\infty$ correspond to no reaction and instant reaction, respectively.  Moreover, we assume $\mathcal{T}_A=\{0\}$ and $\mathcal{T}_B=\{100\}$ $\mu$s and initial concentrations $C_A^{\mathrm{init}}(\mathbf{r})=C_B^{\mathrm{init}}(\mathbf{r})=0$. In Fig.~\ref{Fig:CR_kf}, we show the corresponding \ac{CR} of the type-$A$ molecules, $\bar{y}_A(t)$,  versus time. As can be seen from this figure, for the non-reactive scenario, $\kappa_f=0$, the \ac{CR} is very wide which causes significant \ac{ISI} for the considered nominal value of symbol duration, i.e., $T^{\mathrm{symb}}=200$ $\mu$s. On the other hand, as $\kappa_f$ increases, the \ac{CR} is shortened and correspondingly \ac{ISI} is reduced. For example, for the considered default system parameters, i.e., $\kappa_f=10^{-17}$, $T^{\mathrm{symb}}=200$ $\mu$s, and sampling time $T^{\mathrm{samp}}=100$ $\mu$s after the beginning of the symbol interval, \ac{ISI} is reduced by a factor of $14$ compared to the non-reactive case. Furthermore, the reason why the \ac{CR} $\bar{y}_A(t)$ does not suddenly drop to zero after the release of the type-$B$ molecules when $\kappa_f\to\infty$ is that the type-$B$ molecules need a finite time to diffuse and reach the type-$A$ molecules in the channel.  In this case, the \ac{CR} can be further shortened only by increasing $D_B$, cf. Fig.~\ref{Fig:CR_DB}. Note  that since we assume $\kappa_b=0$, the equilibrium level $\bar{y}_A^{\mathrm{eq}}$ is zero.

\begin{figure}
  \centering  
\resizebox{0.9\linewidth}{!}{\psfragfig{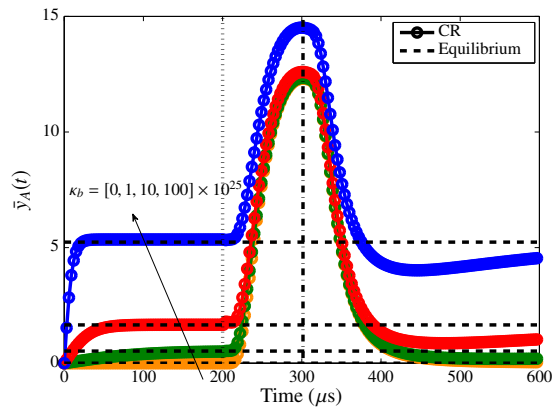}}  \vspace{-0.3cm}
\caption{CR $\bar{y}_A(t)$ for $k_b=[0, 1, 10, 100]\times 10^{25}$ molecule$\cdot$m$^{-3}\cdot$s$^{-1}$ and $k_f=10^{-17}$ molecule$^{-1}\cdot$m$^3\cdot$s$^{-1}$. The vertical dotted and dashed-dotted lines denote the release times of type-$A$ and type-$B$ molecules, respectively.}
\label{Fig:CR_kb}
\end{figure}

Finally, we study the effect of backward reaction rate constant $\kappa_b\in\{0,1,10,100\}\times 10^{25}$ molecule$\cdot$m$^{-3}\\ \cdot$s$^{-1}$ on the \ac{CR} when $\kappa_f=10^{-17}$ molecule$^{-1}\cdot$m$^3\cdot$s$^{-1}$. Moreover, we assume $\mathcal{T}_A=\{200\}$ and $\mathcal{T}_B=\{300\}$ $\mu$s and the initial concentrations are zero. From Fig.~\ref{Fig:CR_kb}, we observe that the main effect of increasing $\kappa_b$ is that the equilibrium level increases as well. This implies that more noise molecules are generated which can impair the performance of the \ac{MC} system, cf. Remark~\ref{Remk:Equilib}. Interestingly, the channel dispersion does not significantly change  with the values of $\kappa_b$. In other words, $\kappa_b$ impacts the level of noise molecules in the \ac{MC} channel and does not considerably affect \ac{ISI}.

In summary, Figs.~\ref{Fig:CR_DB}-\ref{Fig:CR_kb} indicate that  \ac{MC} systems generally perform better by choosing molecules with large diffusion coefficients (for more effective \ac{ISI} reduction as well as faster transmission), large forward reaction rates (for more effective \ac{ISI} reduction), and small backward reaction rates (to decrease the number of noise molecules).

\begin{figure*} 	
	\hspace{2.5cm}
	\begin{minipage}{0.32\textwidth}
		\centering  
		\resizebox{1\linewidth}{!}{\psfragfig{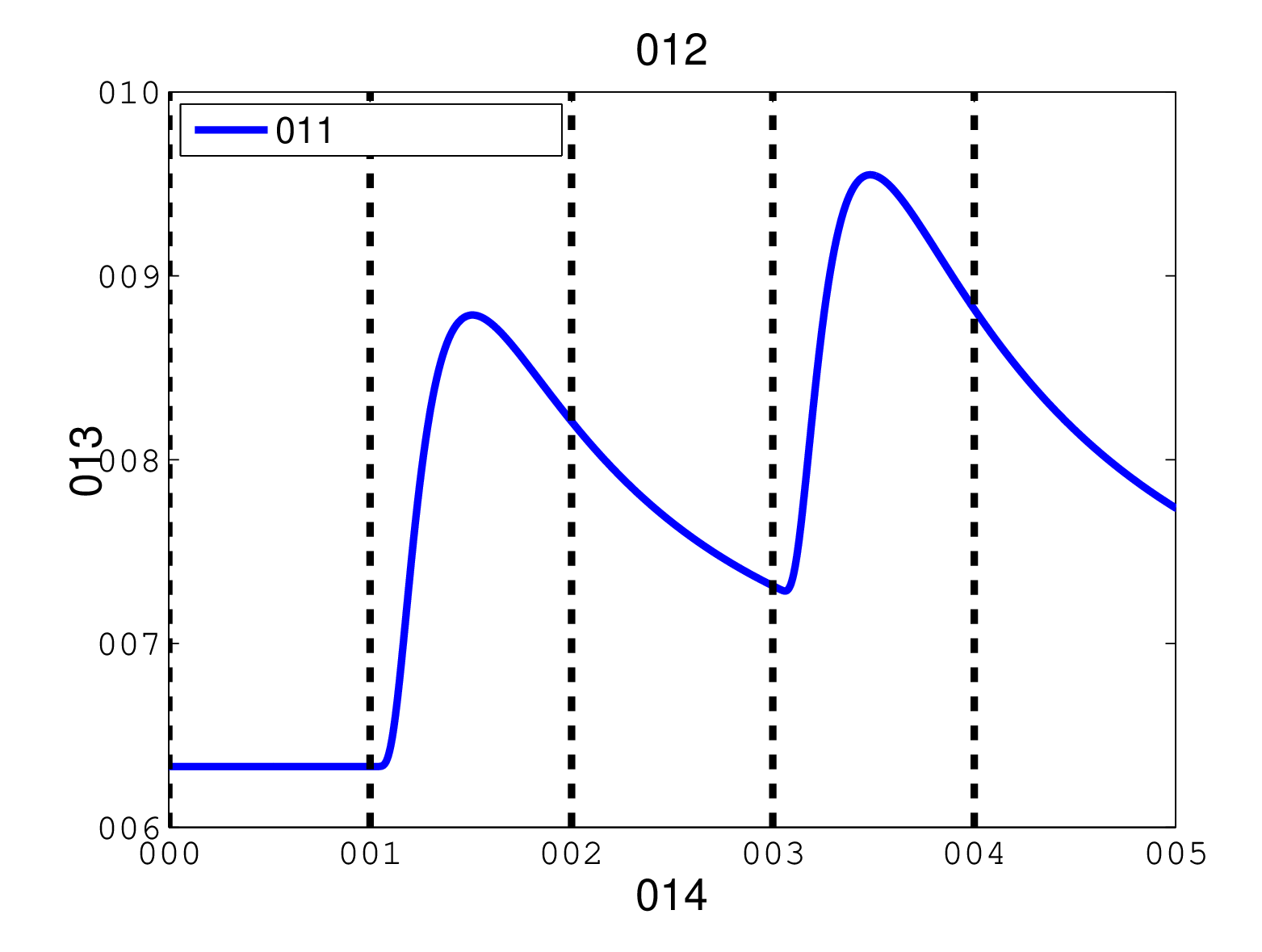}} 
	\end{minipage}
	\begin{minipage}{0.32\textwidth}
		\centering  
		\resizebox{1\linewidth}{!}{\psfragfig{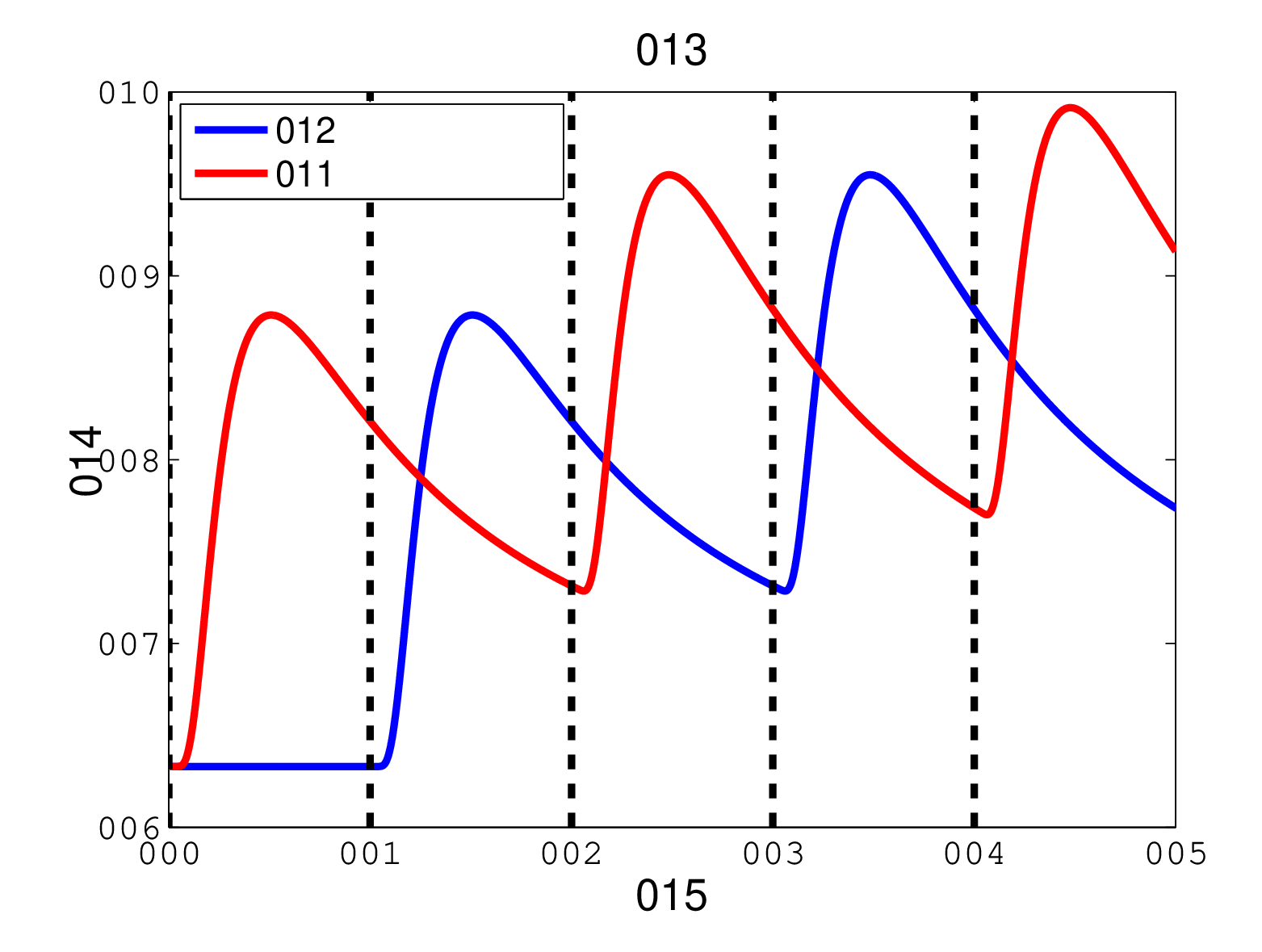}} 
	\end{minipage}
	
	\begin{minipage}{0.32\textwidth}
		\centering  
		\resizebox{1\linewidth}{!}{\psfragfig{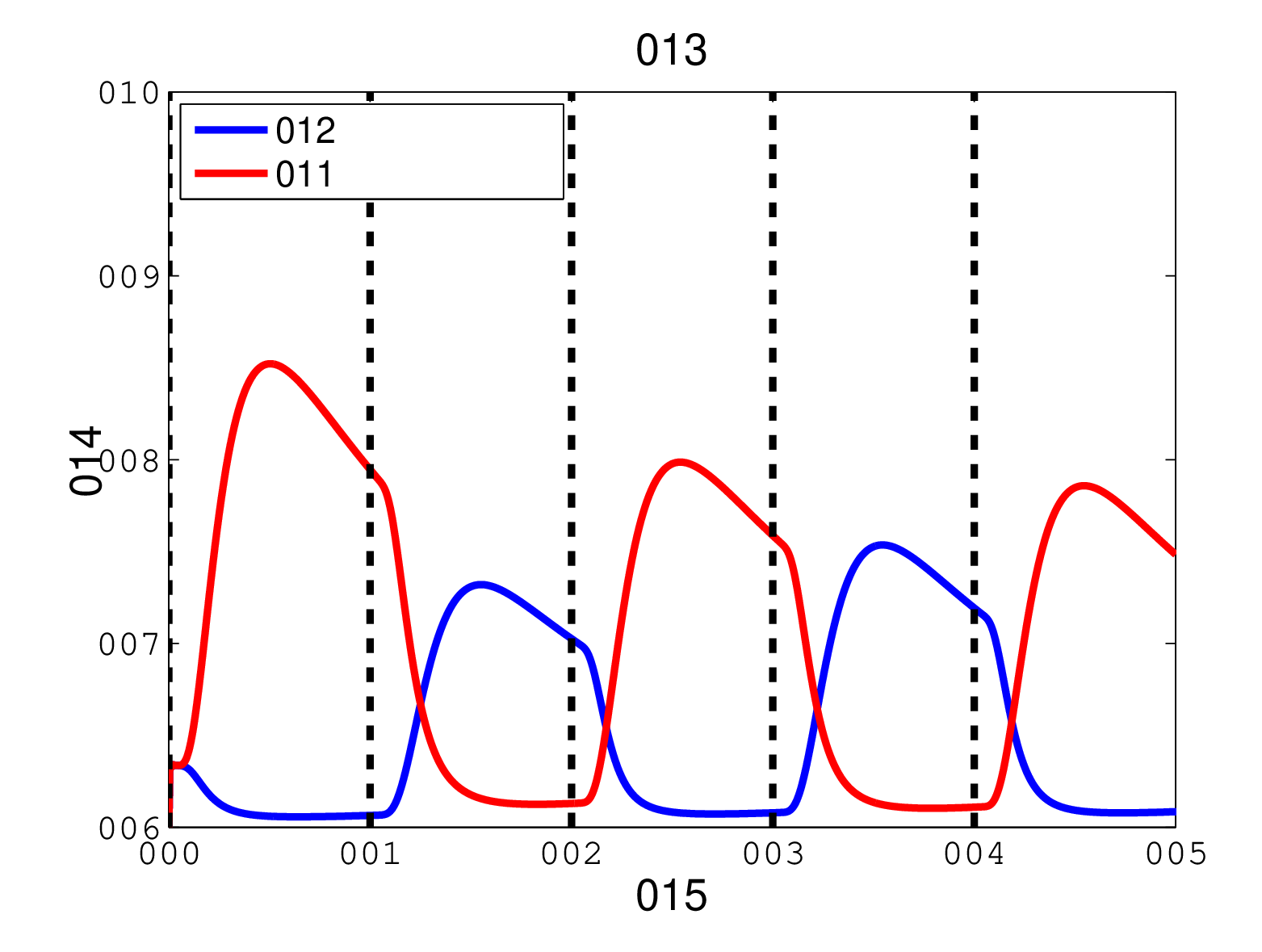}} 
	\end{minipage}
	\begin{minipage}{0.32\textwidth}
		\centering  
		\resizebox{1\linewidth}{!}{\psfragfig{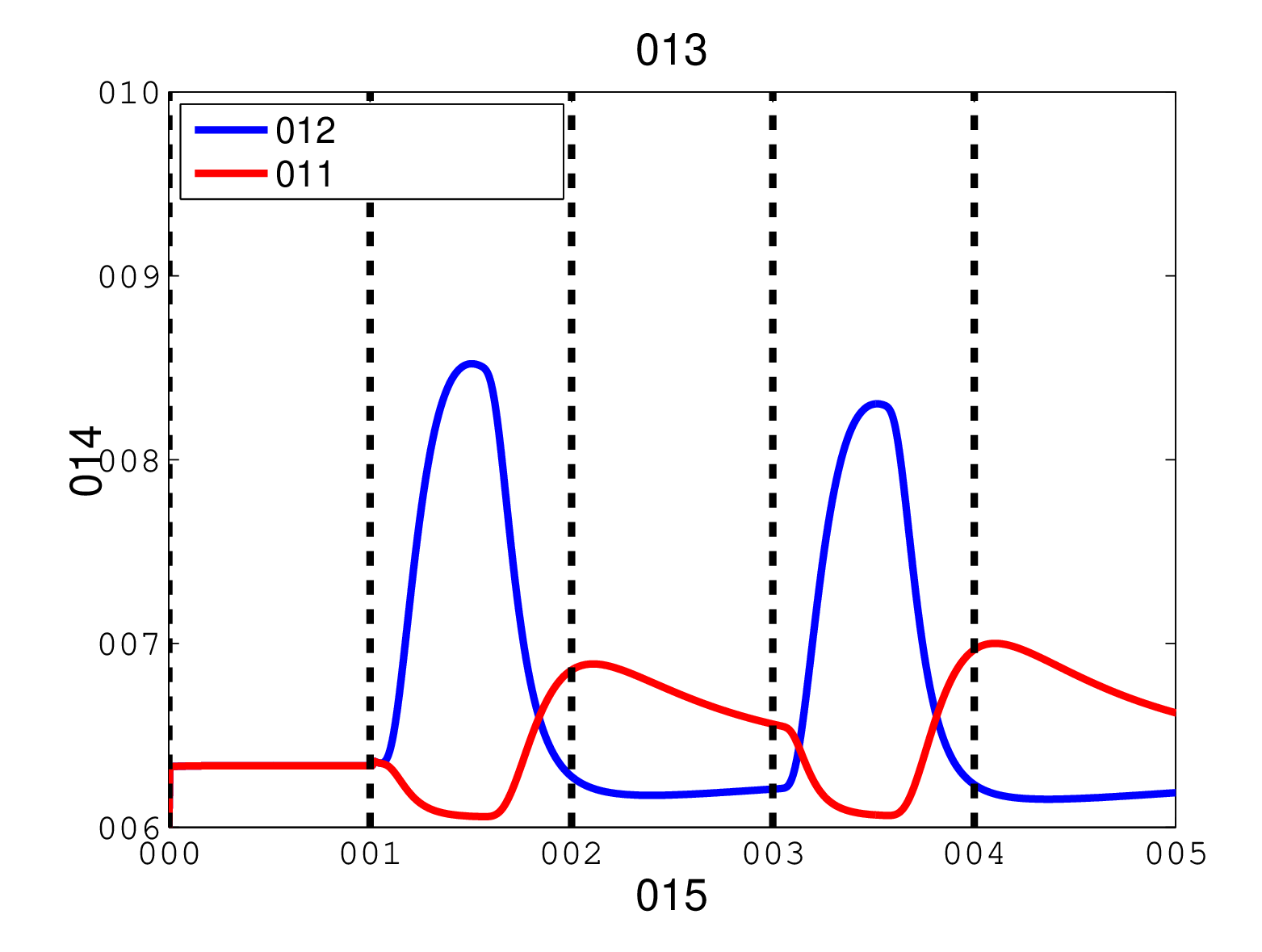}} 
	\end{minipage}
	\begin{minipage}{0.32\textwidth}
		\centering  
		\resizebox{1\linewidth}{!}{\psfragfig{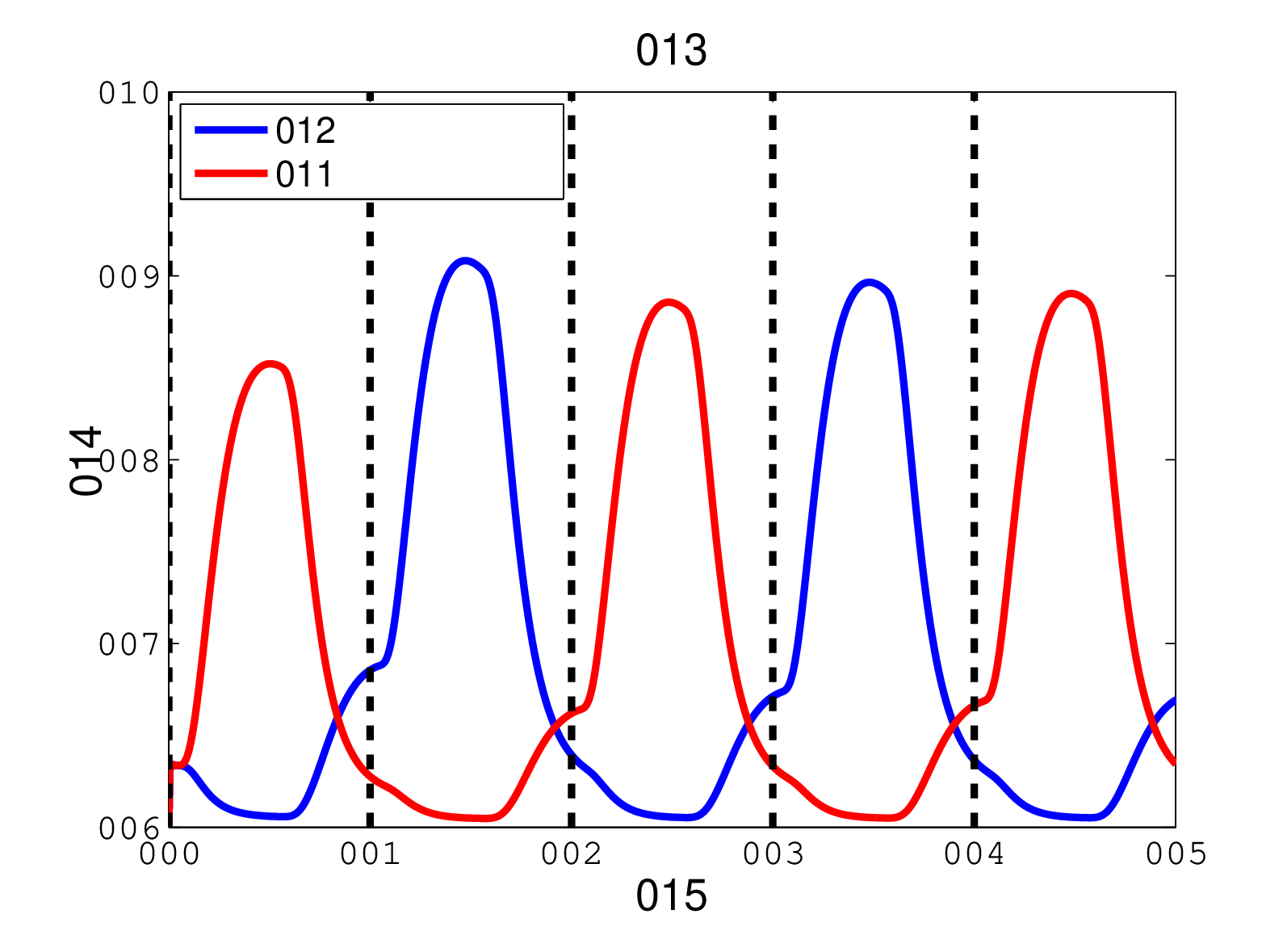}} 
	\end{minipage}
	\caption{CRs $\bar{y}_A(t)$ and $\bar{y}_B(t)$ for sequence $\mathbf{s}=[0,1,0,1,0]$. The vertical dashed lines denote the beginning of a new symbol interval.}
	\label{Fig:CR_Mod}
\end{figure*}

\subsection{Performance Evaluation}
In this section, we evaluate the performance of the modulation and detection schemes proposed in this paper. 
We consider both 1TM and 2TM receivers. As benchmarks for non-reactive MC systems, we consider the following three modulation schemes: \textit{i)} conventional OOK modulation with 1TM receiver \cite{Nariman_Survey}, \textit{ii)} MoSK modulation with 2TM receiver \cite{MoSK_Yilmaz}, and \textit{iii)} the proposed OOK modulation with 2TM receiver where $\tau_1=0$\footnote{When the signaling molecules are not reactive, the proposed OOK modulation scheme reduces to sending the same data over two orthogonal channels using conventional OOK modulation which was previously studied in \cite{tepekule2015novel}. For a fair comparison, unlike \cite{tepekule2015novel} which employs $y(t)=y_A(t)-y_B(t)$ as the received signal, we perform joint ML detection based on $y_A(t)$ and $y_B(t)$. Note that based on the data processing inequality, optimal joint processing of $y_A(t)$ and $y_B(t)$ in general yields a better performance than the processing of any function of $y_A(t)$ and $y_B(t)$ including $y(t)$. Moreover, in the non-reactive case, choosing $\tau_1=0$ yields the best performance since both  $y_A(t)$ and $y_B(t)$ attain their respective peak values at the same time.}.  For reactive \ac{MC} systems, we consider the three modulation schemes introduced in Section~\ref{Sec:Transmitter}, namely conventional MoSK, the proposed OOK, and the proposed OSK.  Note that the optimal and suboptimal detectors derived in Section~\ref{Sec:Analysis} are applicable to all of the above binary modulation schemes in non-reactive and reactive \ac{MC} systems.

In Fig.~\ref{Fig:CR_Mod}, we show CR $\bar{y}_A(t)$ and $\bar{y}_B(t)$ versus time for data sequence $\mathbf{s}=[0,1,0,1,0]$. We assume $\kappa_f=10^{-17}$ molecule$^{-1}\cdot$m$^3\cdot$s$^{-1}$ and $\kappa_b=10^{26}$ molecule$\cdot$m$^{-3}\cdot$s$^{-1}$ which leads to a equilibrium level of $\bar{y}_i^{\mathrm{eq}}=1.65$ for the reactive \ac{MC} system. Since this equilibrium level can be interpreted as the expected number of noise molecules within the receiver volume, cf. Remark~\ref{Remk:Equilib}, we assume that the non-reactive system is affected by the same fixed level of noise molecules. As can be seen from Fig.~\ref{Fig:CR_Mod},  for non-reactive signaling, $\bar{y}_i(t)$ increases over time which is due to the significant \ac{ISI} from previous symbol intervals. On the contrary, for reactive signaling, the peak of $\bar{y}_i(t)$ remains almost constant. This is due to the fact that signaling molecules participate in a degradation reaction and cancel each other out. Moreover, employing reactive signaling molecules provides a huge potential for generating pulse shapes with desirable properties, i.e., causing minimum \ac{ISI}. For the reactive systems, we observe from Fig.~\ref{Fig:CR_Mod} that the CRs of the proposed OOK and OSK modulation schemes are narrower than those of conventional MoSK modulation which results in lower ISI. We emphasize that although, in this paper,  we assumed an unbounded simulation environment, we expect that degradation via reactive molecules becomes even more important in a bounded environment where the accumulation of molecules can significantly contaminate the channel.

Now, we study the performance of the considered modulation and detection schemes in terms of BER. To this end, we simulate $10^5$ realizations of blocks of $K=10$ symbols and use $L=5$ for the proposed \ac{ML} detector. We emphasize that, for simulation of $y_A$ and $y_B$, the memory of all previous symbol intervals is considered, whereas for the \ac{ML} detector, only a memory length of size $L=5$ is assumed for simplicity. 

We first focus on a reactive \ac{MC} system. Fig.~\ref{Fig:BER_sub} shows the BER vs. the number of released molecules, $N^{\mathrm{tx}}_A = N^{\mathrm{tx}}_B\triangleq N^{\mathrm{tx}}$, for the genie-aided (GA) lower bound from Proposition~\ref{Prop:MLDetector} and Corollary~\ref{Corol:MLDetector}, the \ac{ML}-based detector with estimated \ac{ISI}, and the suboptimal ISI-neglecting detector in (\ref{Eq:MLdetector_simpleAB}) and (\ref{Eq:MLdetector_simpleA}). Fig.~\ref{Fig:BER_sub} a) and b) show the results for 1TM and 2TM receivers, respectively. The following observations can be made in Fig.~\ref{Fig:BER_sub}:

\begin{itemize}
\item Let us first focus on the ML-based detector with estimated \ac{ISI}. The \ac{ML}-based detector  performs very close to the corresponding genie-aided lower bound for all considered modulation schemes and both receiver types.  For MoSK and OSK modulation, since information is encoded in both molecule types, they are expected to perform well for 2MT receivers. Fig.~\ref{Fig:BER_sub} supports this expectation. Interestingly, the proposed OSK modulation performs well even if a simple 1TM receiver is used whereas the performance of conventional MoSK significantly deteriorates. On the other hand, since information is encoded in type-$A$ molecules in OOK modulation\footnote{Although the proposed OOK modulation for reactive \ac{MC} systems employs two types of molecules, the information is primarily encoded in the concentration of type-$A$ molecules. In fact, type-$B$ molecules are mainly used for \ac{ISI} reduction, see Fig.~\ref{Fig:CR_Mod}.}, it is expected to perform well for both the 1TM receiver (that observes type-$A$ molecules) and the 2TM receiver. This is also consistent with the results shown in Fig.~\ref{Fig:BER_sub}. Overall,  we observe from Fig.~\ref{Fig:BER_sub} that for all considered modulation schemes, the 2TM receiver outperforms the 1TM receiver by exploiting the diversity gain that the observation of two molecule types offers.  This additional gain of the 2TM receiver over the 1TM receiver is significant for MoSK and OSK modulation but is smaller for OOK modulation.
\begin{figure} 
	\resizebox{0.9\linewidth}{!}{\psfragfig{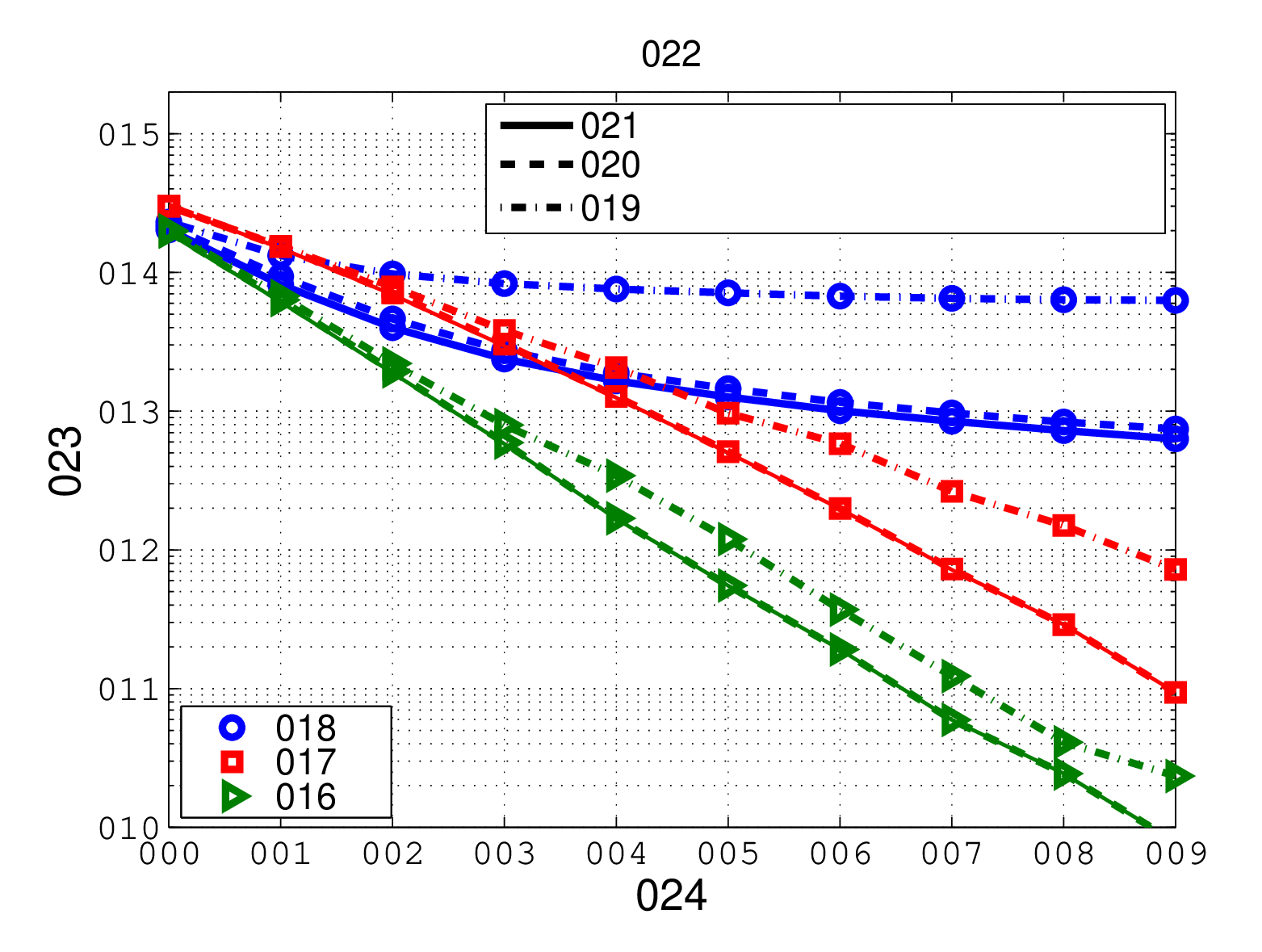}} \vspace{-0.3cm}
	\centering  
	\resizebox{0.9\linewidth}{!}{\psfragfig{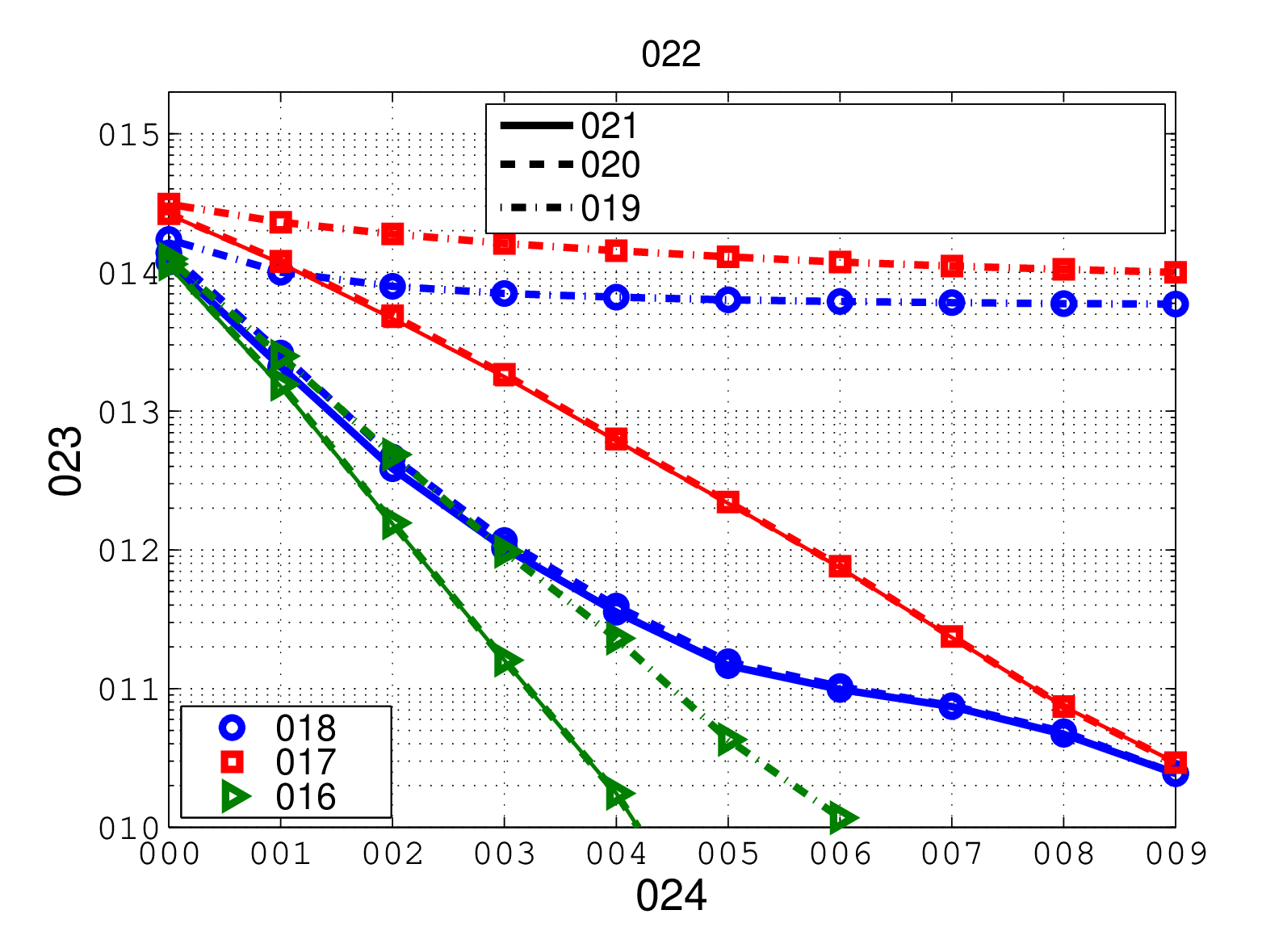}} \vspace{-0.1cm}
	\caption{BER vs. number of released molecules, $N^{\mathrm{tx}}_A = N^{\mathrm{tx}}_B\triangleq N^{\mathrm{tx}}$ for a) 1TM receivers and b) 2TM receivers. }
	\label{Fig:BER_sub}
\end{figure}
\item Now, we focus on the suboptimal ISI-neglecting detector. For the conventional MoSK modulation, the performance of the suboptimal ISI-neglecting detector is not good for both 1TM and 2TM receivers. In fact, for MoSK modulation, the BER curves of the ISI-neglecting detector for both 1TM and 2TM receivers have an error floor at large $N^\mathrm{tx}$. The reason for this behavior is that this suboptimal scheme was proposed for the case when ISI is negligible whereas ISI is not sufficiently suppressed with conventional MoSK modulation. For the proposed OOK modulation, the ISI-neglecting detector performs well with respect to the corresponding genie-aided bound only if a 1TM receiver is used. For the 2TM receiver and OOK modulation, the ISI-neglecting detector in (\ref{Eq:MLdetector_simpleAB}) does not perform well since this detector was designed for a modulation that encodes information in the concentration of two types of molecules which is not the case for OOK modulation. Finally, for the proposed OSK modulation, we observe that the ISI-neglecting detector has good performance relative to the genie-aided bound for both the 1TM and 2TM receivers. In fact, OSK modulation is well suited for 2TM receivers since at the sampling time,  $\Ybxi{A}{1}\gg \Ybxi{B}{1}$ and $\Ybxi{A}{0}\ll \Ybxi{B}{0}$ hold. OSK modulation also performs well for 1TM receivers since at the sampling time, $\Ybxi{A}{1}\gg \Ybxi{A}{0}$ holds, see Fig.~\ref{Fig:CR_Mod}. These conditions hold due to effective ISI reduction for OSK modulation.
\item In summary, among the considered modulation and detection schemes, OSK modulation with the suboptimal ISI-neglecting detector  provides a favorable tradeoff between performance and complexity for both receiver types. Hereby, if a simpler 1TM receiver is adopted (compared to the 2TM receiver), the \ac{CR} has to be acquired to determine the detection threshold $\gamma$. On the other hand, if a more complex 2TM receiver is adopted, the ISI-neglecting detector does not require any knowledge of the  \ac{CR} for its operation.
\end{itemize}

In Fig.~\ref{Fig:BER_comp}, we compare the performance of reactive and non-reactive \ac{MC} systems. In particular, Fig.~\ref{Fig:BER_comp} shows the BER vs. the number of released molecules, $N^{\mathrm{tx}}_A = N^{\mathrm{tx}}_B\triangleq N^{\mathrm{tx}}$, for the ML-based detector. We summarize our observations from Fig.~\ref{Fig:BER_comp} in the following:

\begin{itemize}
\item From Fig.~\ref{Fig:BER_comp}, we see that the \ac{MC} system with reactive signaling molecules has superior performance compared to the \ac{MC} system with non-reactive signaling molecules for all considered modulation schemes and moderate-to-large $N^{\mathrm{tx}}$.  This is due to the fact that  for the adopted symbol duration, \ac{ISI} is significantly reduced with reactive signaling; by contrast, for non-reactive signaling, the \ac{ISI} is severe, particularly for large $N^{\mathrm{tx}}$, and thus limits performance. 
In fact, at large $N^{\mathrm{tx}}$, the slope of the \ac{BER} curve  for the proposed OOK and OSK modulation schemes is larger than that for the benchmark schemes which is due to their effective use of chemical reactions to reduce ISI. 
\item For OOK modulation with both reactive and non-reactive molecules, although the 2TM receiver outperforms the 1TM receiver, the slopes of the BER curves are similar for both receiver types. This behavior can be justified as follows. 
For non-reactive MC systems, the BER curve saturates for large $N^{\mathrm{tx}}$ due to severe ISI which ultimately limits the slope of the BER curves for both 1TM and 2TM receivers (and all considered modulation schemes). For reactive MC systems, type-$A$ molecules are the primary signaling molecules for the proposed OOK modulation and type-$B$ molecules are mainly used for cleaning the channel for the next release of type-$A$ molecules. Therefore, the observation of type-$B$ molecules at the receiver contains little information for the symbol sent in the current interval which explains the marginal gain of the 2TM receiver over the 1TM receiver in this case.
\begin{figure} 
	\centering  
	\resizebox{0.9\linewidth}{!}{\psfragfig{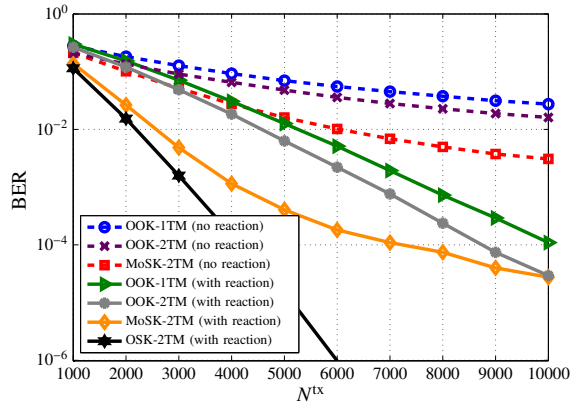}} \vspace{-0.3cm}
	\caption{BER of ML-based detector vs. number of released molecules, $N^{\mathrm{tx}}_A = N^{\mathrm{tx}}_B\triangleq N^{\mathrm{tx}}$. \vspace{-0.5cm} }
	\label{Fig:BER_comp}
\end{figure}
\item For reactive \ac{MC} systems, the slope of the \ac{BER} curve for OSK modulation with 2TM receivers is larger than that of OOK modulation with 1TM and 2TM receivers. This is due to the effective use of two types of molecules in OSK modulation whose observation at the receiver provides an increased diversity gain (the slope of BER curve). 
\item Interestingly, for MoSK modulation in reactive \ac{MC} systems, the slope of the BER curve is higher for small-to-moderate $N^{\mathrm{tx}}$ and decreases for large $N^{\mathrm{tx}}$. The reason for this behavior is attributed to high BER of those sequences that contain multiple consecutive ones or zeros. These sequences introduce a severe ISI which cannot be cope with in reactive MC systems using MoSK modulation.
\end{itemize}
The analytical investigation of the diversity gains of these schemes is an interesting topic for future research.

\begin{figure}
	\centering 
		\resizebox{0.9\linewidth}{!}{\psfragfig{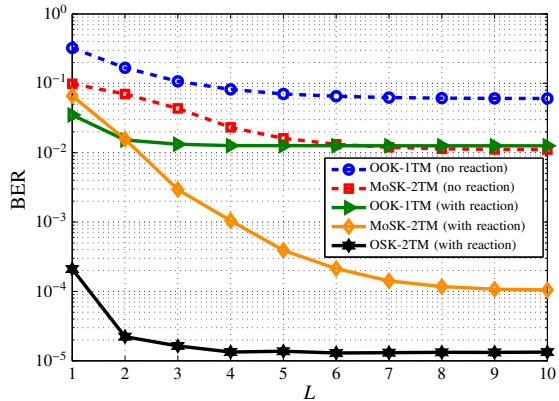}} \vspace{-0.3cm}
		\caption{BER of ML-based detector vs. $L$ for $K=10$ and $N^{\mathrm{tx}}_A = N^{\mathrm{tx}}_B = 5\times 10^{3}$. \vspace{-0.5cm} }
		\label{Fig:BER_L}	
\end{figure}

\begin{figure} 
		\centering  
		\resizebox{0.9\linewidth}{!}{\psfragfig{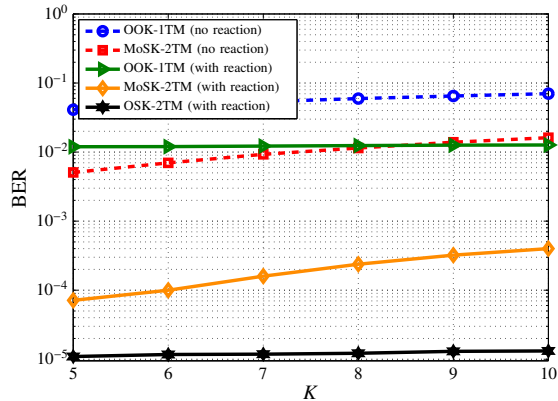}} \vspace{-0.3cm}
		\caption{BER of ML-based detector vs. $K$ for $L=5$ and $N^{\mathrm{tx}}_A = N^{\mathrm{tx}}_B = 5\times 10^{3}$. \vspace{-0.5cm} }
		\label{Fig:BER_K}	
\end{figure}

Recall that for the simulation results presented in Figs.~\ref{Fig:BER_sub} and \ref{Fig:BER_comp}, we assume blocks of length $K=10$ and generate $y_i(\mathbf{s}),\,\,i\in\{A,B\}$, without any CR truncation, i.e., the molecules observed at the tenth symbol interval may originate from the release in the first symbol interval. However, for detection, we take into account a channel memory length of $L=5$ for simplicity. In Figs.~\ref{Fig:BER_L} and \ref{Fig:BER_K}, we study the impact of parameters $L$ and $K$. In particular, in Fig.~\ref{Fig:BER_L}, we plot the BER vs. $L$ for $K=10$ and $N^{\mathrm{tx}}_A = N^{\mathrm{tx}}_B = 5\times 10^{3}$. We observe from this figure that for the proposed OOK and OSK modulations, the ISI is negligible even for $L>3$ due to efficient ISI reduction enabled by chemical reactions. For the benchmark schemes, the impact of ISI is still observable for $L$ between $6$ and $8$ symbols. Overall, Fig.~\ref{Fig:BER_L} implies that, for the proposed modulation schemes, the actual channel memory is smaller than the considered block length of $K=10$. Therefore, we expect that considering larger block lengths does not significantly affect the performance. This is validated in Fig.~\ref{Fig:BER_K} where the BER is plotted vs. block length $K$ for $L=5$ and $N^{\mathrm{tx}}_A = N^{\mathrm{tx}}_B = 5\times 10^{3}$. We observe from this figure  that for the proposed OOK and OSK modulations in reactive MC systems, the BER remains almost constant as $K$ increases, which is due to the efficient ISI reduction inherent to these modulation schemes. For the benchmark schemes, the BER slightly increases as the block length increases. Therefore, considering $K>10$ does not change the BER of the proposed OOK and OSK modulation schemes whereas it may slightly deteriorate the performance of the benchmark schemes.	

\section{Conclusions and Future Work}\label{Sec:Conclusions}

We studied an \ac{MC} system that employs two types of molecules for signaling. These molecules may participate in a reversible bimolecular reaction in the channel. We also considered two receiver types, namely 1TM and 2TM receivers. For this system, we proposed two binary modulation schemes, i.e., a modified OOK and OSK, and derived an optimal genie-aided \ac{ML} detector and two suboptimal detectors, an \ac{ML}-based detector and an \ac{ISI}-neglecting detector. Knowledge of the \ac{CR} was needed for all detectors except the ISI-neglecting detector for the 2TM receiver as well as for performance evaluation. However, deriving the \ac{CR} of a reactive MC system is challenging since the underlying \ac{PDE}s that describe the reaction-diffusion mechanism are coupled and non-linear. To address this issue,  we developed a numerical algorithm for efficient computation of the \ac{CR}. The accuracy of this algorithm has been validated via particle-based simulations. Moreover, simulation results confirmed the superior performance of reactive systems over non-reactive systems  due to efficient \ac{ISI} reduction enabled by chemical reactions. In particular, the BER performance of the proposed OOK and OSK modulations improves as the number of released molecules increases whereas the BERs of conventional OOK and MoSK modulations with non-reactive signaling molecules saturate due to severe ISI. In addition, for all considered modulation schemes and receiver types, the ML-based detectors yielded a similar performance as the corresponding genie-aided detectors. Furthermore, for OSK modulation, we observed that the ISI-neglecting detector provided a favorable tradeoff between performance and complexity and hence is a desirable candidate for practical \ac{MC} systems with limited computational capabilities.  The results of this paper can be further extended to non-binary signaling and other MC environments, e.g., considering other geometries and incorporating flow.

\appendices

\section{Proof of Proposition~\ref{Prop:MLDetector}}\label{App:Prop_MLDetector} %

The \ac{LLR} for  problem (\ref{Eq:MLprobAB})  can be written as
\begin{IEEEeqnarray}{rll} \label{Eq:LLR}
\text{LLR}=&\log\left(\frac{f_{\mathcal{P}}(y_A|s=1,\mathbf{s})f_{\mathcal{P}}(y_B|s=1,\mathbf{s})}{f_{\mathcal{P}}(y_A|s=0,\mathbf{s})f_{\mathcal{P}}(y_B|s=0,\mathbf{s})}\right) \nonumber \\
= &\log\left(\frac{(\bar{y}_A^{(1)}(\mathbf{s}))^{y_A} e^{-\bar{y}_A^{(1)}(\mathbf{s})} (\bar{y}_B^{(1)}(\mathbf{s}))^{y_B} e^{-\bar{y}_B^{(1)}(\mathbf{s})}}{(\bar{y}_A^{(0)}(\mathbf{s}))^{y_A} e^{-\bar{y}_A^{(0)}(\mathbf{s})} (\bar{y}_B^{(0)}(\mathbf{s}))^{y_B} e^{-\bar{y}_B^{(0)}(\mathbf{s})}}\right) \nonumber \\
= &y_A \log\left(\frac{\bar{y}_A^{(1)}(\mathbf{s})}{\bar{y}_A^{(0)}(\mathbf{s})}\right)
-y_B \log\left(\frac{\bar{y}_B^{(0)}(\mathbf{s})}{\bar{y}_B^{(1)}(\mathbf{s})}\right) \nonumber \\ &-\bar{y}_A^{(1)}(\mathbf{s})-\bar{y}_B^{(1)}(\mathbf{s})+\bar{y}_A^{(0)}(\mathbf{s})+\bar{y}_B^{(0)}(\mathbf{s}).
\end{IEEEeqnarray}
Due to the monotonicity of the logarithm, the \ac{ML} problem in (\ref{Eq:MLprobAB}) can be rewritten as $\text{LLR} \overset{s=1}{\underset{s=0}{\gtreqless}} 0$.  Defining $\alpha(\mathbf{s})$ and $\beta(\mathbf{s})$  as in Proposition~\ref{Prop:MLDetector}, we obtain (\ref{Eq:MLdetectorAB}) which concludes the proof.

\section{Proof of Corollary~\ref{Corol:MLDetector}}\label{App:Corol_MLDetector} %

	The proof is similar to that given in Appendix~\ref{App:Prop_MLDetector} for 2TM receivers. In particular, the LLR for detection problem (\ref{Eq:MLprobA}) can be written as
	\begin{IEEEeqnarray}{lll} 
		\text{LLR}&=\log\left(\frac{f_{\mathcal{P}}(y_A|s=1,\mathbf{s})}{f_{\mathcal{P}}(y_A|s=0,\mathbf{s})}\right) = \log\left(\frac{(\bar{y}_A^{(1)}(\mathbf{s}))^{y_A} e^{-\bar{y}_A^{(1)}(\mathbf{s})} }{(\bar{y}_A^{(0)}(\mathbf{s}))^{y_A} e^{-\bar{y}_A^{(0)}(\mathbf{s})}}\right) \nonumber \\
		&= y_A \log\left(\frac{\bar{y}_A^{(1)}(\mathbf{s})}{\bar{y}_A^{(0)}(\mathbf{s})}\right) -\bar{y}_A^{(1)}(\mathbf{s})+\bar{y}_A^{(0)}(\mathbf{s}).
	\end{IEEEeqnarray}
	The ML problem in (8) can be rewritten as $\text{LLR} \overset{s=1}{\underset{s=0}{\gtreqless}} 0$ which in turn can be reformulated into $y_A \overset{s=1}{\underset{s=0}{\gtreqless}} \gamma(\mathbf{s})$ where  $\gamma(\mathbf{s})$ is defined in Corollary~1. This completes the proof.

\section{Proof of Lemma~\ref{Lem:Numerical}}\label{App:Lem_Numerical}

The concentration changes due to diffusion and reaction  given in (\ref{Eq:Reaction_Diff_integral}) can be written in form of
\begin{IEEEeqnarray}{lll} 
\Delta C_i^{\mathrm{x}}(\mathbf{r},t+\Delta t) =  \int_{\tilde{t}=t}^{t+\Delta t} f_i^{\mathrm{x}}(\mathbf{r},\tilde{t}) \mathrm{d}\tilde{t}, \quad \mathrm{x}\in\{\mathrm{df},\mathrm{rc}\},
\end{IEEEeqnarray}
where $f_i^{\mathrm{df}}(\mathbf{r},\tilde{t})=D_i \nabla^2 C_i(\mathbf{r},\tilde{t})$ and $f_i^{\mathrm{rc}}(\mathbf{r},\tilde{t})=\kappa_b- \kappa_f C_A(\mathbf{r},\tilde{t})C_B(\mathbf{r},\tilde{t})$. Let us define $\bar{f}_i^{\mathrm{df}}(\mathbf{r},\tilde{t})=D_i \nabla^2 C_i^{\mathrm{df}}(\mathbf{r},\tilde{t})$ where $C_i^{\mathrm{df}}(\mathbf{r},\tilde{t})$  denotes the concentration of type-$i$ molecules assuming during interval $\tilde{t}\in[t,t+\Delta t]$ diffusion is present and reaction is absent. Similarly, let us define $\bar{f}_i^{\mathrm{rc}}(\mathbf{r},\tilde{t})=\kappa_b- \kappa_f C_A^{\mathrm{rc}}(\mathbf{r},\tilde{t})C_B^{\mathrm{rc}}(\mathbf{r},\tilde{t})$ where $C_i^{\mathrm{rc}}(\mathbf{r},\tilde{t})$  denotes the concentration of type-$i$ molecules assuming during interval $\tilde{t}\in[t,t+\Delta t]$ reaction is present and diffusion is absent. For mathematical rigorousness, we distinguish the following two cases:

\textbf{Case 1:} In this case, we assume that the transmitter releases no molecules within interval $[t,t+\Delta t]$, i.e., $(\mathcal{T}_A\cup\mathcal{T}_B)\cap[t,t+\Delta t]$ is empty. Thereby, by defining $\epsilon_i^{\mathrm{x}}(\mathbf{r},\tilde{t})\triangleq f_i^{\mathrm{x}}(\mathbf{r},\tilde{t})-\bar{f}_i^{\mathrm{x}}(\mathbf{r},\tilde{t})$, we have 
\begin{IEEEeqnarray}{lll} 
\Delta C_i^{\mathrm{x}}(\mathbf{r},t+\Delta t) 
= \underset{\triangleq A_i^{\mathrm{x}}(\Delta t)}{\underbrace{\int_{\tilde{t}=t}^{t+\Delta t} \bar{f}_i^{\mathrm{x}}(\mathbf{r},\tilde{t})\mathrm{d}\tilde{t}}} 
+\underset{\triangleq B_i^{\mathrm{x}}(\Delta t)}{\underbrace{\int_{\tilde{t}=t}^{t+\Delta t} \epsilon_i^{\mathrm{x}}(\mathbf{r},\tilde{t}) \mathrm{d}\tilde{t}}}.
\end{IEEEeqnarray}
Next, we show that as $\Delta t \to 0$, $B_i^{\mathrm{x}}(\Delta t)$ becomes negligible compared to $A_i^{\mathrm{x}}(\Delta t)$. Under assumption A1, i.e., $C_i(\mathbf{r},t)\neq 0$ and $\nabla^2 C_i(\mathbf{r},t)\neq 0$, we obtain $\bar{f}_i^{\mathrm{x}}(\mathbf{r},t)\neq 0$ which leads to $\underset{\Delta t \to 0}{\lim} A_i^{\mathrm{x}}(\Delta t)\sim o(\Delta t)$. Recall that $\epsilon_i^{\mathrm{x}}(\mathbf{r},\tilde{t})$ characterizes the error in decoupling the reaction and diffusion phenomena at time $\tilde{t}>t$. Therefore, for $B_i^{\mathrm{x}}(\Delta t)$, we have $\epsilon_i^{\mathrm{x}}(\mathbf{r},t)=0$ which leads to $\underset{\Delta t \to 0}{\lim} B_i^{\mathrm{x}}(\Delta t)\sim o(\Delta t^{p})$ for some $p>1$. In summary, we have $\underset{\Delta t \to 0}{\lim}\frac{B_i^{\mathrm{x}}(\Delta t)}{A_i^{\mathrm{x}}(\Delta t)}\sim o(\Delta t^{p-1})\to 0$.

\textbf{Case 2:} In this case, we assume the transmitter releases molecules within interval $[t,t+\Delta t]$, i.e., $(\mathcal{T}_A\cup\mathcal{T}_B)\cap[t,t+\Delta t]$ is non-empty. For this case, the conditions on set $\mathcal{T}_i$ introduced in assumption A2 are needed such that the argument in Case~1 is also applicable to Case~2.  In particular, under condition $\mathcal{T}_i\subset\mathcal{T}$, without ambiguity, each release by the transmitter is taken into account only in one of the time slots.  Moreover, due to condition $\varepsilon\ll\Delta t$, the impact of diffusion and reaction on molecule concentration in interval $[t+\Delta t - \varepsilon,t+\Delta t]$ is negligible compared to that in $[t,t+\Delta t - \varepsilon)$. Furthermore, assuming that $\mathcal{T}_A\cap\mathcal{T}_B$ is empty is a reasonable assumption since simultaneous release of both types of molecules leads to their degradation via the forward reaction. These conditions enable us to directly apply the proof for $\underset{\Delta t \to 0}{\lim}\frac{B_i^{\mathrm{x}}(\Delta t)}{A_i^{\mathrm{x}}(\Delta t)}\sim o(\Delta t^{p-1})\to 0$ provided for Case~1 to Case~2.  

In summary, under assumptions A1 and A2, we can conclude that as $\Delta t \to 0$, the relative effect arising from the coupling of diffusion and reaction becomes negligible, and the overall concentration change is simply a superposition of the concentration changes due to the individual phenomena. This concludes the proof. 

\section{Proof of Proposition~\ref{Prop:Numerical}}\label{App:Prop_Numerical}

Eq. (\ref{Eq:UpdateRule}) is obtained by applying (\ref{Eq:Separation}) from Lemma~\ref{Lem:Numerical} in (\ref{Eq:Reaction_Diff_integral}). Under assumption A1, we directly obtain  $\bar{G}_i(\mathbf{r})\triangleq \int_{\tilde{t}=t}^{t+\Delta t} G_i(\mathbf{r},t+\Delta t)\mathrm{d}\tilde{t}= \sum_{t_i\in\mathcal{T}_i} \Nxtx{i}  \delta_{\mathrm{d}}(\mathbf{r})\delta_{\mathrm{k}}(t+\Delta t-\varepsilon-t_i)$. In the following, we derive expressions for $C_i^{\mathrm{df}}(\mathbf{r},t+\Delta t)$ and $C_i^{\mathrm{rc}}(\mathbf{r},t+\Delta t)$ as a function of initial conditions at time $t$, respectively. Without loss of generality and in order to simplify the notations, we assume the initial time as zero and obtain the concentrations for $t>0$,  denoted by $C_i^{\mathrm{df}}(\mathbf{r},t)$ and $C_i^{\mathrm{rc}}(\mathbf{r},t)$. 

\textit{Diffusion:} 
The following diffusion-only \ac{PDE}  has to be solved for $t>0$ and arbitrary initial conditions  $C_i(\mathbf{r},0)$ at initial time $t=0$
\begin{IEEEeqnarray}{lll} 
\frac{\partial C_i^{\mathrm{df}}(\mathbf{r},t)}{\partial t} &= D_i \nabla^2 C_i^{\mathrm{df}}(\mathbf{r},t).
\end{IEEEeqnarray}
For one-dimensional diffusion, this problem is solved in \cite[Chapter 1.7]{NonlinearPDE_Debnath} by transforming it into the Laplace domain with respect to space variable $\mathbf{r}$. The generalization of this result to the $n\in\{2,3\}$ dimensional environment given in (\ref{Eq:DiffRctSol}a) is straightforward. Therefore, we skip the derivation due to space constraints and refer the readers to \cite[Chapter 1.7]{NonlinearPDE_Debnath}.

\textit{Reversible Bimolecular Reaction:} 
The \ac{PDE}  for the reaction-only case is given by
\begin{IEEEeqnarray}{lll} \label{Eq:Cons_reaction}
\frac{\partial C_A^{\mathrm{rc}}(\mathbf{r},t)}{\partial t} &= -\kappa_f C_A^{\mathrm{rc}}(\mathbf{r},t)C_B^{\mathrm{rc}}(\mathbf{r},t) +\kappa_b \IEEEyesnumber \IEEEyessubnumber \\
\frac{\partial C_B^{\mathrm{rc}}(\mathbf{r},t)}{\partial t} &= -\kappa_f C_A^{\mathrm{rc}}(\mathbf{r},t)C_B^{\mathrm{rc}}(\mathbf{r},t)  +\kappa_b. \IEEEyessubnumber
\end{IEEEeqnarray}
Subtracting (\ref{Eq:Cons_reaction}b) from (\ref{Eq:Cons_reaction}a), we obtain
\begin{IEEEeqnarray}{lll} 
\frac{\partial \big( C_A^{\mathrm{rc}}(\mathbf{r},t)-C_B^{\mathrm{rc}}(\mathbf{r},t) \big) }{\partial t} = 0,
\end{IEEEeqnarray}
which has the solution $C_A^{\mathrm{rc}}(\mathbf{r},t)-C_B^{\mathrm{rc}}(\mathbf{r},t)=c_1(\mathbf{r})$ where $c_1(\mathbf{r})$ is a constant with respect to $t$ and is set to $c_1(\mathbf{r})=C_A^{\mathrm{rc}}(\mathbf{r},0)-C_B^{\mathrm{rc}}(\mathbf{r},0)$ to satisfy the initial conditions at $t=0$. Substituting $C_B^{\mathrm{rc}}(\mathbf{r},t)=C_A^{\mathrm{rc}}(\mathbf{r},t)-c_1(\mathbf{r})$ into (\ref{Eq:Cons_reaction}a) leads to
\begin{IEEEeqnarray}{lll}\label{Eq:Riccati} 
\frac{\partial C_A^{\mathrm{rc}}(\mathbf{r},t)}{\partial t} = - \left(\kappa_f\left(C_A^{\mathrm{rc}}(\mathbf{r},t)\right)^2-\kappa_f c_1(\mathbf{r})C_A^{\mathrm{rc}}(\mathbf{r},t)-\kappa_b\right).
\end{IEEEeqnarray}
This equation is in form of a Bernoulli equation (or also in form of a Riccati equation) which can be solved by rewriting (\ref{Eq:Riccati}) as 
\begin{IEEEeqnarray}{lll} 
\frac{\partial C_A^{\mathrm{rc}}(\mathbf{r},t)}{\kappa_f\left(C_A^{\mathrm{rc}}(\mathbf{r},t)\right)^2-\kappa_f c_1(\mathbf{r})C_A^{\mathrm{rc}}(\mathbf{r},t)-\kappa_b}
= -\partial t.
\end{IEEEeqnarray}
Integrating both sides of the above equation and using \cite{TableIntegSerie}
\begin{IEEEeqnarray}{lll} 
\int \frac{\mathrm{d}x}{ax^2+bx+c}=\frac{1}{\Delta}\log\left(\frac{\Delta-2ax-b}{\Delta+2ax+b}\right),
\end{IEEEeqnarray}
where $\Delta=\sqrt{b^2-4ac}$, we obtain
\begin{IEEEeqnarray}{lll} 
- t + \tilde{c}_4(\mathbf{r})=
\frac{1}{c_2(\mathbf{r})}\log\left(\frac{c_2(\mathbf{r})-2\kappa_f C_A^{\mathrm{rc}}(\mathbf{r},t)+\kappa_fc_1(\mathbf{r})}
{c_2(\mathbf{r})+2\kappa_f C_A^{\mathrm{rc}}(\mathbf{r},t)-\kappa_fc_1(\mathbf{r})}\right)
\end{IEEEeqnarray}
where $c_2(\mathbf{r})=\sqrt{\kappa_f^2c_1^2(\mathbf{r})+4\kappa_f\kappa_b}$ and $\tilde{c}_4(\mathbf{r})$ is a constant with respect to time. Rewriting the above equation leads to
\begin{IEEEeqnarray}{lll} 
C_A^{\mathrm{rc}}(\mathbf{r},t) = \\
\frac{c_2(\mathbf{r})+\kappa_fc_1(\mathbf{r})-\big(c_2(\mathbf{r})-\kappa_fc_1(\mathbf{r})\big)
\exp\big(-c_2(\mathbf{r})(t-\tilde{c}_4(\mathbf{r}))\big)}
{2\kappa_f\left[1+\exp\big(-c_2(\mathbf{r})(t-\tilde{c}_4(\mathbf{r}))\big)\right]}.\nonumber
\end{IEEEeqnarray}
Using the initial condition $C_A^{\mathrm{rc}}(\mathbf{r},0)$, we obtain $\tilde{c}_4(\mathbf{r})=\frac{1}{c_2(\mathbf{r})}\mathrm{ln}\left(\frac{c_2(\mathbf{r})+\kappa_fc_3(\mathbf{r})}{c_2(\mathbf{r})-\kappa_fc_3(\mathbf{r})}\right)$ with $c_3(\mathbf{r})= C_A^{\mathrm{rc}}(\mathbf{r},0)+C_B^{\mathrm{rc}}(\mathbf{r},0)$. This leads to $C_A^{\mathrm{rc}}(\mathbf{r},t)$ given in (\ref{Eq:DiffRctSol}b) after defining $c_4(\mathbf{r})\triangleq\exp(c_2(\mathbf{r})\tilde{c}_4(\mathbf{r}))=\frac{c_2(\mathbf{r})-\kappa_fc_3(\mathbf{r})}{c_2(\mathbf{r})+\kappa_fc_3(\mathbf{r})}$ for notational simplicity. Using $C_B^{\mathrm{rc}}(\mathbf{r},t)=C_A^{\mathrm{rc}}(\mathbf{r},t)-c_1(\mathbf{r})$, we straightforwardly obtain $C_B^{\mathrm{rc}}(\mathbf{r},t)$ given in (\ref{Eq:DiffRctSol}c).  

The above results are concisely given in Proposition~\ref{Prop:Numerical} which concludes the proof.

\section{Proof of Corollary~\ref{Corol:OneD}}\label{App:Corol_OneD} %

Because of the geometrical symmetry of the problem, the concentration for the \ac{MC} system under consideration  is only a function of  $r\triangleq \|\mathbf{r}\|$. In the following, we present the proof for one-, two-, and three-dimensional environments, respectively. 

\textit{One-Dimensional Environment:} In this case, we obtain (\ref{Eq:Concen_OneD}) by substituting $n=1$ in (\ref{Eq:DiffRctSol}a) as
\begin{IEEEeqnarray}{lll} 
C_i^{\mathrm{df}}(\mathbf{r},t+\Delta t)  \\
 =  \frac{1}{(4\pi D_i \Delta t)^{1/2}} 
\int_{\tilde{r}_1}
C_i(\tilde{r}_1,t) \exp\left(-\frac{|\tilde{r}_1|^2+r^2-2r\tilde{r}_1}{4D_i \Delta t}\right)\mathrm{d}\tilde{r}_1, \nonumber \\
 = \frac{1}{(4\pi D_i \Delta t)^{1/2}} 
\int_{\tilde{r}_1=-\infty}^{0}
C_i(\tilde{r}_1,t) \exp\left(-\frac{|\tilde{r}_1|^2+r^2-2r\tilde{r}_1}{4D_i \Delta t}\right)\mathrm{d}\tilde{r}_1, \nonumber \\
\quad + \frac{1}{(4\pi D_i \Delta t)^{1/2}}  \int_{\tilde{r}_1=0}^{\infty}
C_i(\tilde{r}_1,t) \exp\left(-\frac{|\tilde{r}_1|^2+r^2-2r\tilde{r}_1}{4D_i \Delta t}\right)\mathrm{d}\tilde{r}_1, \nonumber \\
\overset{(a)}{=} \frac{1}{(4\pi D_i \Delta t)^{1/2}} 
\int_{\tilde{r}=0}^{\infty}
C_i(\tilde{r},t) 2\exp\left(-\frac{\tilde{r}^2+r^2}{4D_i \Delta t}\right)
\cosh\left(\frac{\tilde{r}r}{2D_i \Delta t}\right)\mathrm{d}\tilde{r}, \nonumber
\end{IEEEeqnarray}
where for equality $(a)$, we used $\tilde{r}=|\tilde{r}_1|$. Defining $W_i(r,\tilde{r})$ as in (\ref{Eq:WeightFunc}) leads to  (\ref{Eq:Concen_OneD}) for $n=1$.

\textit{Two-Dimensional Environment:} For $n=2$, the simplification of $C_i^{\mathrm{df}}(\mathbf{r},t+\Delta t)$ requires transforming an integral from  Cartesian coordinates to polar coordinates, i.e., 
\begin{IEEEeqnarray}{lll} \label{Eq:CartPolar}
\int_{r_1}\int_{r_2} f(r_1,r_2) \mathrm{d}r_1\mathrm{d}r_2 
= \int_{r}\int_{\theta} f\big(r\cos(\theta),r\sin(\theta)\big)  r  \mathrm{d}r \mathrm{d}\theta, \quad\,\,\,\,
\end{IEEEeqnarray}
where $f(r_1,r_2)$ is an arbitrary function and $r\geq 0$ and $\theta\in[0,2\pi]$ are the variables of the polar coordinates. Moreover, without loss of generality, we consider $\mathbf{r}=(r,0)$ in order to simplify (\ref{Eq:DiffRctSol}a) as follows
\begin{IEEEeqnarray}{lll} 
C_i^{\mathrm{df}}(\mathbf{r},t+\Delta t) \nonumber \\
 =  \frac{1}{4\pi D_i \Delta t} 
\iint_{\tilde{\mathbf{r}}}
C_i(\tilde{\mathbf{r}},t) \exp\left(-\frac{\|\tilde{\mathbf{r}}\|^2+r^2-2r\tilde{r}_1}{4D_i \Delta t}\right)\mathrm{d}\tilde{\mathbf{r}}, \nonumber \\
 = \frac{1}{4\pi D_i \Delta t} 
\int_{\tilde{r}=0}^{\infty} C_i(\tilde{r},t) 
\tilde{r}\exp\left(-\frac{\tilde{r}^2+r^2}{4D_i \Delta t}\right) \nonumber \\
 \qquad\qquad \times \int_{\tilde{\theta}=0}^{2\pi}\exp\left(\frac{r\tilde{r}\cos(\tilde{\theta})}{2D_i \Delta t}\right) 
\mathrm{d}\tilde{\theta} \mathrm{d}\tilde{r}, \quad\,\,\, \nonumber \\
 \overset{(a)}{=} \frac{1}{4\pi D_i \Delta t}
\int_{\tilde{r}=0}^{\infty} C_i(\tilde{r},t) 
2\pi \tilde{r}
\exp\left(-\frac{\tilde{r}^2+r^2}{4D_i \Delta t}\right)
I_0\left(\frac{r\tilde{r}}{2D_i \Delta t}\right)  \mathrm{d}\tilde{r}, \quad\,\,\,
\end{IEEEeqnarray}
where in equality $(a)$,  we used the identity $\int_{x=0}^{2\pi} \exp(c\cos(x))\mathrm{d}x = 2\pi I_0(c)$ \cite{TableIntegSerie}. Defining $W_i(r,\tilde{r})$ as in (\ref{Eq:WeightFunc}) leads to  (\ref{Eq:Concen_OneD}) for $n=2$. 

\textit{Three-Dimensional Environment:} In a similar manner, the simplification of $C_i^{\mathrm{df}}(\mathbf{r},t+\Delta t)$ when  $n=3$ requires transforming an integral from  Cartesian coordinates to spherical coordinates, i.e., 
\begin{IEEEeqnarray}{lll} \label{Eq:CartSphr}
\int_{r_1}\int_{r_2}\int_{r_3} f(r_1,r_2,r_3) \mathrm{d}r_1\mathrm{d}r_2\mathrm{d}r_3 \nonumber \\
= \int_{r}\int_{\phi}\int_{\theta} f\big(r\sin(\phi)\cos(\theta),r\sin(\phi)\sin(\theta),r\cos(\phi)\big) \nonumber \\
\qquad\qquad \times r^2 \sin(\phi) \mathrm{d}r
\mathrm{d}\phi \mathrm{d}\theta, \quad 
\end{IEEEeqnarray}
where $f(r_1,r_2,r_3)$ is an arbitrary function and $r\geq 0$, $\phi\in[0,\pi]$, and $\theta\in[0,2\pi]$ are the variables of the spherical coordinates. Here, without loss of generality, we consider $\mathbf{r}=(0,0,r)$ in order to simplify (\ref{Eq:DiffRctSol}a) as follows
\begin{IEEEeqnarray}{lll} 
 C_i^{\mathrm{df}}(\mathbf{r},t+\Delta t)  \nonumber\\
=  \frac{1}{(4\pi D_i \Delta t)^{3/2}} 
\iiint_{\tilde{\mathbf{r}}}
C_i(\tilde{\mathbf{r}},t) \exp\left(-\frac{\|\tilde{\mathbf{r}}\|^2+r^2-2r\tilde{r}_3}{4D_i \Delta t}\right)\mathrm{d}\tilde{\mathbf{r}}, \nonumber \\
 = \frac{1}{(4\pi D_i \Delta t)^{3/2}} 
\int_{\tilde{r}=0}^{\infty} C_i(\tilde{r},t) 
\tilde{r}^2\exp\left(-\frac{\tilde{r}^2+r^2}{4D_i \Delta t}\right) \nonumber \\
\quad \times \int_{\tilde{\phi}=0}^{\pi} \int_{\tilde{\theta}=0}^{2\pi}
\sin(\tilde{\phi})\exp\left(\frac{r\tilde{r}\cos(\tilde{\phi})}{2D_i \Delta t}\right) 
\mathrm{d}\tilde{\theta} \mathrm{d}\tilde{\phi}\mathrm{d}\tilde{r}, \quad\,\,\, \nonumber \\
 \overset{(a)}{=} \frac{1}{(4\pi D_i \Delta t)^{3/2}} 
\int_{\tilde{r}=0}^{\infty} C_i(\tilde{r},t) 
\frac{8\pi  D_i \tilde{r}\Delta t}{r} \nonumber\\
\quad\times\exp\left(-\frac{\tilde{r}^2+r^2}{4D_i \Delta t}\right)
\sinh\left(\frac{r\tilde{r}}{2D_i\Delta t}\right) \mathrm{d}\tilde{r}, 
\end{IEEEeqnarray}
where for equality $(a)$, we used the identity $\int_{x=0}^{\pi} \sin(x)\exp(c\cos(x))\mathrm{d}x = \frac{2\sinh(c)}{c}$. Defining $W_i(r,\tilde{r})$ as in (\ref{Eq:WeightFunc}) leads to  (\ref{Eq:Concen_OneD}) for $n=3$ and completes the proof.

\iftoggle{arXiv}{%
		
	\section{Stochastic Simulation}\label{Sec:SimParticle}
	Stochastic simulation is used to verify the accuracy of Algorithm~\ref{Alg:CR} for \ac{CR} computation and the Poisson statistics in (\ref{Eq:Poisson_ISI}). We refer the readers to \cite{ReactionDiffSim,Erban2009stochastic,Adam_AcCoRD,jamali2018channel} for an overview of general stochastic simulation of reaction-diffusion systems. In the following, we explain the particular particle-based simulator that is used for performance verification in this paper.
	
	For particle-based simulation, it is assumed that the spatial orientations and internal energy levels of the molecules fluctuate on time scales that are faster than the diffusion and reaction processes that are of interest \cite{ReactionDiffSim}. Therefore, it suffices  to track only the positions of the individual molecules  during the simulation time. Let $\mathbf{r}_i(t)=(r_{1,i}^{(m)}(t),r_{2,i}^{(m)}(t),r_{3,i}^{(m)}(t))$ denote the coordinate of a specific type-$i$ molecule, indexed by $m$, at time $t$. Moreover, let $M_i$ denote a variable which tracks the total number of type-$i$ molecules generated in the simulation environment. In the following, we explain how we update the position of the molecules for the considered release, diffusion, and reaction mechanisms. For convenience of presentation, we focus on a three-dimensional environment.
	
	\subsection{Transmitter Release}
	For instantaneous release from a point-source transmitter, we simply place $N^{\mathrm{tx}}_i$ type-$i$ molecules at the origin, i.e., $\mathbf{r}_i^{(m)}(t) = (0,0,0),\,\,m=M_i+1,\dots,M_i+N^{\mathrm{tx}}_i$, at any release time instant $t\in\mathcal{T}_i$. Variable $M_i$ is updated by $M_i+N^{\mathrm{tx}}_i$ after each release of type-$A$ molecules by the transmitter.
	
	\subsection{Diffusion} 
	According to Brownian dynamics, the position of each molecule at time $t+\Delta t$ is updated as \cite{ReactionDiffSim}
	\begin{IEEEeqnarray}{lll} \label{Eq:Brownian}
		\mathbf{r}_i^{(m)}(t+\Delta t) = \mathbf{r}_i^{(m)}(t) + \sqrt{2D_i\Delta t} \big(\Delta r_{1,i}^{(m)},\Delta r_{2,i}^{(m)},\Delta r_{3,i}^{(m)}),
	\end{IEEEeqnarray}
	where $\Delta r_{1,i}^{(m)},\Delta r_{2,i}^{(m)},\Delta r_{3,i}^{(m)}\sim\mathcal{N}(0,1)$ and $\mathcal{N}(\mu,\sigma^2)$ denotes a normal \ac{RV} with mean $\mu$ and variance~$\sigma^2$.
	
	\subsection{Forward Reaction}
	
	The forward reaction in (\ref{Eq:Reaction}), i.e., $ A+B\overset{\kappa_f}{\rightarrow} \varnothing$, is a second order bimolecular reaction.
	One approach for stochastic simulation of these reactions is based on the fact that a reaction occurs within interval $[t, t+\Delta t]$ when the reactant molecules are within a certain binding radius $\rho_b$ \cite{ReactionDiffSim}.  Therefore, in our simulation, we can simply remove one type-$A$ and one type-$B$ molecule if their distance is less than $\rho_b$. In other words, we remove each pair of molecules whose position variables $\mathbf{r}_A^{(m)}(t)$ and $\mathbf{r}_B^{(m')}(t)$ satisfy $\|\mathbf{r}_A^{(m)}(t)-\mathbf{r}_B^{(m')}(t)\|<\rho_b$. Unfortunately, the exact value of $\rho_b$ cannot be found analytically in general  and depends on reaction rate constant $\kappa_f$ and the choice of $\Delta t$. Nevertheless, for the two special cases of $\Delta t\to 0$ and $\Delta t\to \infty$, the following simple relations are available \cite[Eqs. (19) and (20)]{ReactionDiffSim}
	\begin{IEEEeqnarray}{lll} \label{Eq:BindingRadius}
		\kappa_f = \begin{cases}
			4\pi\rho_b(D_A+D_B),\quad &\mathrm{if}\,\,\rho_{\mathrm{rms}}\ll \rho_b \\
			4\pi\rho_b^3/(3\Delta t),\quad &\mathrm{if}\,\,\rho_{\mathrm{rms}} \gg \rho_b
		\end{cases}
	\end{IEEEeqnarray}
	where $\rho_{\mathrm{rms}}=\sqrt{2(D_A+D_B)\Delta t}$ is the mutual root mean square step length of type-$A$ and type-$B$ molecules. Using the simplified formula for $\rho_b$ in (\ref{Eq:BindingRadius}), we obtain equivalent conditions for $\rho_{\mathrm{rms}} \ll \rho_b$ and $\rho_{\mathrm{rms}} \gg \rho_b$ as
	\begin{IEEEeqnarray}{lll} \label{Eq:BindingRadius_Time}
		\begin{cases}
			\rho_{\mathrm{rms}}\ll \rho_b \rightarrow \Delta t \ll T^{\mathrm{crt}}\\
			\rho_{\mathrm{rms}} \gg \rho_b \rightarrow \Delta t \gg \frac{9}{4} T^{\mathrm{crt}},
		\end{cases}
	\end{IEEEeqnarray}
	where $T^{\mathrm{crt}}=\frac{\kappa_f^2}{32\pi^2(D_A+D_B)^3}$. In order to be able to use the closed-form relation in (\ref{Eq:BindingRadius}), we have to choose either large $\Delta t \gg \frac{9}{4} T^{\mathrm{crt}}$ or small $\Delta t \ll T^{\mathrm{crt}}$. This choice depends on the system under consideration and the total length of the simulation interval.

	\subsection{Backward Reaction}
	The backward reaction in  (\ref{Eq:Reaction}), i.e., $\varnothing \overset{\kappa_b}{\rightarrow} A + B$, is in form of a zeroth order reaction. Suppose that the simulation environment is a cube of volume $V= L^3$. Moreover, let \ac{RV} $N(t)$ denote the number of times that the backward reaction occurs in a time interval $[t,t+\Delta t]$. Then, $N(t)$ follows a Poisson distribution \cite{ReactionDiffSim}
	\begin{IEEEeqnarray}{lll} \label{Eq:ZeroOrderPoisson}
		N(t) = \mathcal{P}\big(V\kappa_b\Delta t\big).
	\end{IEEEeqnarray}
	Here, we have to be careful about the positions of  the type-$A$ and type-$B$ molecules that are generated via the backward reaction. In particular, if $\rho_{\mathrm{rms}} \ll \rho_b$ holds and we put the  molecules generated by the backward reaction at the same location, then these molecules will directly participate in the forward reaction before they can diffuse away  regardless of the value of $\rho_b$ \cite{ReactionDiffSim,Adam_AcCoRD}.  In order to avoid the automatic degradation of  type-$A$ and type-$B$ molecules, the type-$A$ and type-$B$ product molecules are initially separated by a fixed distance, referred to as the unbinding radius $\rho_u$, which is larger than $\rho_b$ .
	Let  $\mathbf{l}$ denote the center of the cube that defines the simulation environment. Then,  the zeroth order reaction can be simulated by inserting each of the $N(t)$ molecules of type-$i$ at uniformly random positions $\mathbf{r}_i^{(m)}(t),\,\,m=M_i+1,\dots,M_i+N(t)$,  obtained from \cite{ReactionDiffSim}
	\begin{IEEEeqnarray}{lll} \label{Eq:ZeroOrderPosition}
		\mathbf{r}_A^{(m)}(t) &= \mathbf{l} +  L\big(\Delta r_{1,A}^{(m)},\Delta r_{2,A}^{(m)},\Delta r_{3,A}^{(m)}) \IEEEyesnumber\IEEEyessubnumber \\
		\mathbf{r}_B^{(m)}(t) &= \mathbf{r}_A^{(m)}(t) + \rho_u\big(\Delta r_{1,B}^{(m)},\Delta r_{2,B}^{(m)},\Delta r_{3,B}^{(m)} \big),\IEEEyessubnumber
	\end{IEEEeqnarray}
	where $\Delta r_{1,A}^{(m)},\Delta r_{2,A}^{(m)},\Delta r_{3,A}^{(m)}\sim\mathcal{U}(-0.5,0.5)$ and $\mathcal{U}(a,b)$ is an \ac{RV} uniformly distributed in the interval $[a,b]$. Moreover, we have $\Delta r_{1,B}^{(m)}=\sin(\theta^{(m)})\cos(\phi^{(m)}),\Delta r_{2,B}^{(m)}=\sin(\theta^{(m)})\sin(\phi^{(m)})$, and $\Delta r_{3,B}^{(m)}=\cos(\theta^{(m)})$ where $\theta^{(m)}\sim\mathcal{U}(0,\pi)$ and $\phi^{(m)}\sim\mathcal{U}(0,2\pi)$. On the other hand, if $\rho_{\mathrm{rms}} \gg \rho_b$ holds and diffusion is simulated after the backward reaction in the adopted simulator, it is very likely that the molecules diffuse out of the binding radius after one diffusion step. In this case, the value of the unbinding radius is not important and  without loss of generality, we can choose $\rho_u=0$ which leads to $\mathbf{r}_A^{(m)}(t) = \mathbf{r}_B^{(m)}(t)$. Variable $M_i$ is updated by $M_i+N(t)$ at the end of this step. 
	
	Algorithm~\ref{Alg:Particle} summarizes the main steps required for particle-based simulation of the considered \ac{MC} system.
	
	\begin{algorithm}[t]
		\caption{Particle-based Simulation}
		\begin{algorithmic}[1]\label{Alg:Particle}
			\STATE \textbf{initialize:} $t=0$, $\Delta t$, $T^{\max}$, $\mathbf{r}_i^{(m)}(0)$, $M_i$, and $\mathcal{T}_i,\,\,i\in\{A,B\}$.   
			\WHILE{$t\leq T^{\max}$}
			\IF{$t\in\mathcal{T}_A$}
			\STATE \textit{Transmitter input:}  Add $N^{\mathrm{tx}}_A$ type-$A$ molecules at position $\mathbf{r}_A^{(m)}(t) = (0,0,0),\,\,m=M_A+1,\dots,M_A+N^{\mathrm{tx}}_A$ and update $M_A$ by $M_A+N^{\mathrm{tx}}_A$.  
			\ELSIF{$t\in\mathcal{T}_B$}
			\STATE \textit{Transmitter input:}  Add $N^{\mathrm{tx}}_B$ type-$B$ molecules at position $\mathbf{r}_B^{(m)}(t) = (0,0,0),\,\,m=M_B+1,\dots,M_B+N^{\mathrm{tx}}_B$ and update $M_B$ by $M_B+N^{\mathrm{tx}}_B$.   
			\ENDIF
			\STATE \textit{Diffusion:} Update the positions of molecules $\mathbf{r}_i^{(m)}(t),\,\,\forall m,i$, based on~(\ref{Eq:Brownian}).
			\STATE \textit{Degradation:} Remove any pair of type-$A$ and type-$B$ molecules whose positions satisfy $\|\mathbf{r}_A^{(m)}-\mathbf{r}_B^{(m')}\|\leq\rho_b$.
			\STATE \textit{Production:} Compute $N(t)$ from (\ref{Eq:ZeroOrderPoisson}), add $N(t)$ type-$i,\,\,i\in\{A,B\},$ molecules at positions $\mathbf{r}_i^{(m)}(t),\,\,m=M_i+1,\dots,M_i+N(t)$ given in (\ref{Eq:ZeroOrderPosition}), and update $M_i$ by $M_i+N(t)$.
			\STATE Assign $y_i(t),\,\,i\in\{A,B\}$, as the number  of type-$i$ molecules whose positions satisfy $\mathbf{r}_i^{(m)}\in\mathcal{V}^{\mathrm{rx}}$.
			\STATE Update $t$ with  $t+\Delta t$.
			\ENDWHILE
			\STATE Return $y_A(t)$ and $y_B(t)$.
		\end{algorithmic}
	\end{algorithm}
	
}{%
}

\bibliographystyle{IEEEtran}
\bibliography{Paper}

\iftoggle{arXiv}{%
}{%
\begin{IEEEbiography}[{\includegraphics[width=1in,height=1.25in,clip,keepaspectratio]{Fig/Author/Vahid}}]{Vahid Jamali} (S'12) received the B.S. and M.S. degrees (honors) in electrical engineering from the
	K. N. Toosi University of Technology, Tehran,
	Iran, in 2010 and 2012, respectively, and
	the Ph.D. degree (with distinctions) from
	the Friedrich-Alexander-University (FAU) of
	Erlangen-N\"urnberg, Erlangen, Germany,
	in 2019. In 2017, he was a Visiting Research Scholar with Stanford
	University, CA, USA. He is currently a Postdoctoral Fellow
	with the Institute for Digital Communication, FAU. His 
	research interests include wireless and molecular communications,
	Bayesian inference and learning, and multiuser information~theory.
	
	Dr. Jamali has served as a member of the Technical Program
	Committee for several IEEE conferences. He received several
	awards, including the Exemplary Reviewer Certificates from \textsc{IEEE Communications Letters}
	in 2014 and the \textsc{IEEE Transactions on Communications} in 2017 and 2018, the Best Paper Award from
	the IEEE International Conference on Communications in 2016,
	the Doctoral Scholarship from the German Academic Exchange
	Service (DAAD) in 2017, the Goldener Igel Publication Award from
	the Telecommunications Laboratory (LNT), FAU, in 2018, and
	he was the Winner of the Best 3 Minutes Ph.D. Thesis (3MT)
	Presentation from the IEEE Wireless Communications and Networking Conference in 2018. He is currently an Associate Editor of \textsc{IEEE Communications Letters}.
\end{IEEEbiography}

\begin{IEEEbiography}[{\includegraphics[width=1in,height=1.25in,clip,keepaspectratio]{Fig/Author/Nariman}}]{Nariman Farsad} (S’07-M'15) is currently a Postdoctoral Fellow with the Department of Electrical Engineering at Stanford University, where he is a recipient of Natural Sciences and Engineering Research Council of Canada (NSERC) Postdoctoral Fellowship. His research interests are on emerging communication technologies such as molecular communication, and on improving the performance of communication systems through machine learning and deep learning. 
	
Dr. Farsad has won the second prize in 2014 IEEE ComSoc Student Competition: Communications Technology Changing the World, the best demo award at INFOCOM’2015, and was recognized as a finalist for the 2014 Bell Labs Prize. He has been an Area Associate Editor for \textsc{IEEE Journal of Selected Areas of Communication}--Special Issue on Emerging Technologies in Communications, and a Technical Reviewer for a number of journals including IEEE Transactions on Signal Processing, and IEEE Transactions on Information Theory. He was also a member of the Technical Program Committees for the ICC’2015, ICC’2018, BICT’2015, GLOBCOM’2015, GLOBCOM’2016, and GLOBECOM’2017. 
\end{IEEEbiography}

\begin{IEEEbiography}[{\includegraphics[width=1in,height=1.25in,clip,keepaspectratio]{Fig/Author/Robert}}]{Robert Schober} (S'98, M'01, SM'08, F'10) was born in Neuendettelsau, Germany, in 1971. He received the Diplom (Univ.) and the Ph.D. degrees in electrical engineering from the Friedrich-Alexander-University of Erlangen-Nuremberg (FAU), Germany, in 1997 and 2000, respectively. From May 2001 to April 2002 he was a Postdoctoral Fellow at the University of Toronto, Canada, sponsored by the German Academic Exchange Service (DAAD). From 2002 to 2011, he was a Professor and Canada Research Chair at the University of British Columbia (UBC), Vancouver, Canada. Since January 2012 he is an Alexander von Humboldt Professor and the Chair for Digital Communication at FAU. His research interests fall into the broad areas of Communication Theory, Wireless Communications, and Statistical Signal Processing.
	
	Dr. Schober received several awards for his work including the 2002 Heinz Maier–Leibnitz Award of the German Science Foundation (DFG), the 2004 Innovations Award of the Vodafone Foundation for Research in Mobile Communications, the 2006 UBC Killam Research Prize, the 2007 Wilhelm Friedrich Bessel Research Award of the Alexander von Humboldt Foundation, the 2008 Charles McDowell Award for Excellence in Research from UBC, a 2011 Alexander von Humboldt Professorship, a 2012 NSERC E.W.R. Stacie Fellowship, and a 2017 Wireless Communications Recognition Award. Furthermore, he has been listed as a Highly Cited Researcher by Clarivate Analytics. Robert is a Fellow of the Canadian Academy of Engineering and a Fellow of the Engineering Institute of Canada. From 2012 to 2015, he served as Editor-in-Chief of the \textsc{IEEE Transactions on Communications}. Currently, he is the Chair of the Steering Committee of the \textsc{IEEE Transactions on Molecular, Biological and Multiscale Communication} and serves on the Editorial Board of the Proceedings of the IEEE. Furthermore, he is a Member at Large of the Board of Governors and a Distinguished Lecturer of the IEEE Communications Society.
\end{IEEEbiography}

\begin{IEEEbiography}[{\includegraphics[width=1in,height=1.25in,clip,keepaspectratio]{Fig/Author/Andrea}}]{Andrea Goldsmith} (S'90-M'93-SM'99-F'05) received
	the B.S., M.S., and Ph.D. degrees in electrical
	engineering from the University of California at
	Berkeley. She is the Stephen Harris Professor in the School
	of Engineering and a Professor of Electrical Engineering
	at Stanford University. She was previously
	on the faculty of Electrical Engineering at Caltech.
	Her research interests are in information theory
	and communication theory, and their application to
	wireless communications and related fields. She cofounded
	and served as Chief Scientist of Plume WiFi, and previously cofounded
	and served as CTO of Quantenna Communications, Inc. She has also
	held industry positions at Maxim Technologies, Memorylink Corporation, and
	AT\&T Bell Laboratories. She is author of the book Wireless Communications
	and co-author of the books MIMO Wireless Communications and Principles of
	Cognitive Radio, all published by Cambridge University Press, as well as an
	inventor~on~28~patents.
	
	Dr. Goldsmith has received several awards for her work, including the
	IEEE ComSoc Edwin H. Armstrong Achievement Award, as well as Technical
	Achievement Awards in Communications Theory and in Wireless Communications,
	the National Academy of Engineering Gilbreth Lecture Award,
	the IEEE ComSoc and Information Theory Society Joint Paper Award, the
	IEEE ComSoc Best Tutorial Paper Award, the Alfred P. Sloan Fellowship,
	the WICE Technical Achievement Award, and the Silicon Valley/San Jose
	Business Journals Women of Influence Award. She has served on the Steering
	Committee for the \textsc{IEEE Transactions on Wireless Communications}
	and as editor for the \textsc{IEEE Transactions on Information Theory},
	the Journal on Foundations and Trends in Communications and Information
	Theory and in Networks, the \textsc{IEEE Transactions on Communications},
	and the IEEE Wireless Communications Magazine. She participates actively
	in committees and conference organization for the IEEE Information Theory
	and Communications Societies and has served on the Board of Governors for
	both societies. She has also been a Distinguished Lecturer for both societies,
	served as President of the IEEE Information Theory Society in 2009, founded
	and chaired the student committee of the IEEE Information Theory society,
	and chaired the Emerging Technology Committee of the IEEE Communications
	Society. At Stanford she received the inaugural University Postdoc Mentoring
	Award, served as Chair of Stanford’s Faculty Senate and for multiple terms as a
	Senator, and currently serves on its Budget Group, Committee on Research, and
	Task Force on Women and Leadership. She is a Fellow of Stanford~University.
\end{IEEEbiography}
}

\end{document}